\pdfoutput=1
\documentclass[11pt,twoside,a4paper,cmspaper,final,collab]{cms-tdr}
\def\svnVersion{b512ca1}\def\svnDate{2024/07/05}\def\cmsCernNoTag{CERN-EP-2023-297}\def\cmsCernDate{\today}\def\cmsMessage{Published in Reports on Progress in Physics as \href{http://dx.doi.org/10.1088/1361-6633/ad4e65}{\doi{10.1088/1361-6633/ad4e65}.}}
\begin{document}\cmsNoteHeader{BPH-22-005}

\renewcommand{\PKstz}{\ensuremath{\PK^{\ast}(892)^{0}}\xspace} 
\newcommand\PKstp{\ensuremath{\PK^{\ast}(892)^{+}}\xspace} 
\newcommand\splot{\ensuremath{{}_{s}\mathcal{P}\text{lot}}\xspace}
\newcommand\BKll{\ensuremath{\PBp \!\to\! \PKp\ell^+\ell^-}\xspace}
\newcommand\BKSll{\ensuremath{\PBz \!\to\! \PKstz\ell^+\ell^-}\xspace}
\newcommand\BKmm{\ensuremath{\PBp \!\to\! \PKp\MM}\xspace}
\newcommand\BKee{\ensuremath{\PBp \!\to\! \PKp\EE}\xspace}
\newcommand\BpmKmm{\ensuremath{\PBpm \!\to\! \PKpm\MM}\xspace}
\newcommand\BpmKee{\ensuremath{\PBpm \!\to\! \PKpm\EE}\xspace}
\newcommand\RK{\ensuremath{R(\PK)}\xspace}
\newcommand\iRK{\ensuremath{R(\PK)^{-1}}\xspace}
\newcommand\Rpsi{\ensuremath{R_{\PGyP{2S}}}\xspace}
\newcommand\RJPsi{\ensuremath{R_{\JPsi}}\xspace}
\newcommand\BKJp{\ensuremath{\PBp \!\to\! \JPsi\PKp}\xspace}
\newcommand\BpmKJp{\ensuremath{\PBpm \!\to\! \JPsi\PKpm}\xspace}
\newcommand\BKJpee{\ensuremath{\PBp \!\to\! \JPsi(\EE)\PKp}\xspace}
\newcommand\BKJpmm{\ensuremath{\PBp \!\to\! \JPsi(\MM)\PKp}\xspace}
\newcommand\BKPsimm{\ensuremath{\PBp \!\to\! \PGyP{2S}(\MM)\PKp}\xspace}
\newcommand\BKPsiee{\ensuremath{\PBp \!\to\! \PGyP{2S}(\EE)\PKp}\xspace}
\newcommand\BKJpll{\ensuremath{\PBp \!\to\! \JPsi(\ell^+\ell^-)\PKp}\xspace}
\newcommand\BKPsill{\ensuremath{\PBp \!\to\! \PGyP{2S}(\ell^+\ell^-)\PKp}\xspace}
\newcommand\qsq{\ensuremath{q^2}\xspace}
\newcommand\qsqmin{\ensuremath{q^2_{\text{min}}}\xspace}
\newcommand\qsqmax{\ensuremath{q^2_{\text{max}}}\xspace}
\newcommand\bsll{\ensuremath{\PAQb \!\to\! \PAQs\ell^+\ell^-}\xspace}
\newcommand\bsmm{\ensuremath{\PAQb \!\to\! \PAQs\MM}\xspace}
\newcommand\BKstzJp{\ensuremath{\PBz \!\to\! \JPsi\PKstz}\xspace}
\newcommand\BKstJpmm{\ensuremath{\PBp \!\to\! \JPsi(\MM)\PKstp}\xspace}
\newcommand\BKstpmJp{\ensuremath{\PBp \!\to\! \JPsi\PKstp}\xspace}
\newcommand\BKPsi{\ensuremath{\PBp \!\to\! \PGyP{2S}\PKp}\xspace}
\newcommand\BR{\ensuremath{\mathcal{B}}\xspace}
\newcommand\RBR{\ensuremath{R_{\BR}}\xspace}
\newcommand\ipsig{\ensuremath{\text{IP}_{\text{sig}}}\xspace}
\newcommand\lumunit{\ensuremath{[10^{34}\unit{cm}^{-2}\unit{s}^{-1}]}\xspace}
\newcommand\mmm{\ensuremath{m_{\MM}}\xspace}
\newcommand\mKmm{\ensuremath{m_{\PKp\MM}}\xspace}
\newcommand\mKee{\ensuremath{m_{\PKp\EE}}\xspace}
\newcommand\mKll{\ensuremath{m_{\PKp\ell\ell}}\xspace}
\newcommand\anyB{\ensuremath{\PB^{\ast 0/+}}\xspace}
\newcommand\Kstar{\ensuremath{\PK^{\ast}(892)^{0/+}}\xspace}
\newcommand\ptm{\ensuremath{p^\PGm_{\mathrm{T}}}\xspace}
\newcommand\dthreeD{\ensuremath{\abs{d_{\text{3D}}(\PKp,\, \EE)}}\xspace}
\newcommand\cosa{\ensuremath{\cos\alpha_{\text{2D}}(\PBp)}\xspace}
\newcommand\Aeps{\ensuremath{\mathcal{A}\epsilon}\xspace}
\newcommand\etrig{\ensuremath{\epsilon_{\text{trig}}}\xspace}
\newcommand\Aepst{\ensuremath{\mathcal{A}\epsilon\etrig}\xspace}
\newcommand\intlum{\ensuremath{\int\!\!\!\;{\mathcal L}{\mathrm d}t}\xspace}
\newcommand\mystrut{\rule{0pt}{0.9\normalbaselineskip}}
\newlength\cmsTabSkip\setlength{\cmsTabSkip}{1ex}

\cmsNoteHeader{BPH-22-005}
\title{Test of lepton flavor universality in \texorpdfstring{\BpmKmm}{B+/- to K+/- mu+ mu-} and  \texorpdfstring{\BpmKee}{B+/- to K+/- e+ e-} decays in proton-proton collisions at \texorpdfstring{$\sqrt{s} = 13\TeV$}{sqrt(s) = 13 TeV}}
\titlerunning{Test of lepton flavor universality in \PB to \PK decays at 13\TeV}

\date{\today}

\abstract{
A test of lepton flavor universality in \BpmKmm and \BpmKee decays, as well as a measurement of differential and integrated branching fractions of a nonresonant \BpmKmm decay are presented. The analysis is made possible by a dedicated data set of proton-proton collisions at $\sqrt{s} = 13$\TeV recorded in 2018, by the CMS experiment at the LHC, using a special high-rate data stream designed for collecting about 10 billion unbiased \PQb hadron decays. The ratio of the branching fractions $\BR(\BpmKmm)$ to $\BR(\BpmKee)$ is determined from the measured double ratio \RK of these decays to the respective branching fractions of the \BpmKJp with $\JPsi\!\to\!\MM$ and $\EE$ decays, which allow for significant cancellation of systematic uncertainties. The ratio \RK is measured in the range $1.1 <\qsq < 6.0\GeV^2$, where $q$ is the invariant mass of the lepton pair, and is found to be $\RK=0.78^{+0.47}_{-0.23}$, in agreement with the standard model expectation $\RK \approx 1$. This measurement is limited by the statistical precision of the electron channel. The integrated branching fraction in the same \qsq range, $\BR(\BpmKmm) =  (12.42 \pm 0.68)\times 10^{-8}$, is consistent with the present world-average value and has a comparable precision.}

\hypersetup{
pdfauthor={CMS Collaboration},%
pdftitle={Test of lepton flavor universality in B+/- to K+/- mu+ mu- and B+/- to K+/- e+ e- decays in proton-proton collisions at sqrt(s) = 13 TeV},%
pdfsubject={CMS},%
pdfkeywords={CMS, B physics, lepton flavor universality, B meson, b to sll transitions}}

\maketitle 

\section{Introduction\label{sec:intro}}
In the standard model (SM) of particle physics~\cite{Glashow:1961tr,Weinberg:1967tq,Salam:1968rm}, the charged leptons (electrons \Pe, muons \PGm, and \PGt leptons) have identical couplings to the gauge bosons and thus exhibit a similar behavior, up to the kinematic differences related to their different masses. This is commonly known as lepton flavor universality (LFU). Several tests of LFU have been performed in \PW and \PZ boson decays, which are generally found to be in excellent agreement with the SM predictions~\cite{Workman:2022ynf}. The only hint of possible LFU violation (LFUV), at 2.7 standard deviations $(\sigma)$ in the decay of \PW bosons to \PGt vs.\ light leptons from the CERN LEP era~\cite{ALEPH:2013dgf}, was ruled out by the ATLAS~\cite{ATLAS:2020xea} and CMS~\cite{CMS:2022mhs} experiments at the CERN LHC.

\begin{figure*}[htbp]
    \centering
    \includegraphics[width=0.8\textwidth]{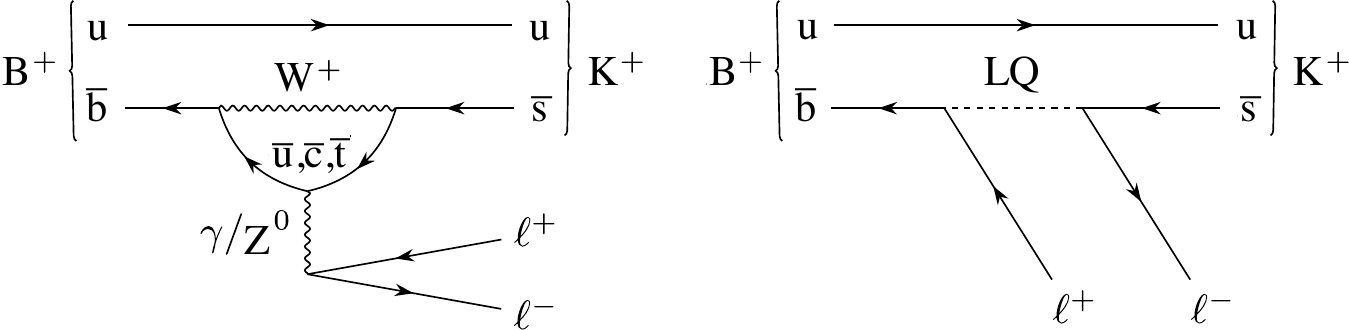} 
    \caption{Representative Feynman diagrams for the decay of a \PBp meson into a \PKp meson and a lepton pair in the SM (left) and in a BSM scenario that introduces a leptoquark (LQ) with flavor-dependent couplings (right).}
    \label{fig:Feynman}
\end{figure*}

Rare \PQb hadron decays provide an excellent and complementary environment to test LFU\@. In particular, the \BKll process where a bottom antiquark (\PAQb) decays into a strange antiquark (\PAQs) and a lepton ($\ell = \PGm$ or $\Pe$) pair is forbidden at tree level and only proceeds via loop diagrams, \eg, the one shown in Fig.~\ref{fig:Feynman} (left). Therefore, the SM branching fractions (\BR) for these decays are very small ($\sim$10$^{-7}$)~\cite{Workman:2022ynf}. This process is referred to as a \bsll transition (in what follows, the charge-conjugated states are implied unless explicitly stated otherwise).

Since both the electron and muon masses are negligible with respect to the \PB meson mass, the available phase space for the two decays (\BKmm and \BKee) is the same to an excellent approximation, which makes the ratio of these branching fractions very close to unity in the SM\@.  On the other hand, beyond-the-SM (BSM) physics processes could modify the corresponding branching fractions differently for different lepton species, as illustrated in Fig.~\ref{fig:Feynman} (right) for the case of a leptoquark with flavor-dependent couplings, thus resulting in LFUV in \bsll transitions. A recent review of various BSM models describing LFUV in \PB meson decays can be found in Ref.~\cite{London:2021lfn} and references therein, as well as in an extensive list of references in Ref.~\cite{LHCb:2021trn}.

A number of tests of LFUV have been performed at the \PB\ and charm factories~\cite{CLEO:2006uhx,BaBar:2010esv,BaBar:2012mrf,BESIII:2018ccy,Belle:2019oag,BELLE:2019xld,Belle:2021dgc,BESIII:2023vfi,BESIII:2024jlj}, as well as at the LHC. 
Over the past decade, the LHCb experiment has reported mounting evidence for LFUV in \BKll~\cite{LHCb:2014vgu,LHCb:2019hip,LHCb:2021trn}, \BKSll~\cite{LHCb:2017avl}, $\PBz \!\to\! \PKzS\ell^+\ell^-$, and $\PBp \!\to\! \PKstp\ell^+\ell^-$~\cite{LHCb:2021lvy} decays with the significance reaching 3.1$\sigma$ in the first channel~\cite{LHCb:2021trn}. In these analyses the muon decays are measured to be suppressed compared to the electron ones. In addition, multiple measurements of branching fractions of several \bsmm decays by LHCb indicate their suppression with respect to the available SM predictions~\cite{LHCb:2012bin,LHCb:2014cxe,LHCb:2016ykl,LHCb:2021zwz}. These state-of-the-art predictions reflect a significant recent theoretical progress in understanding of the \bsll transitions. Nevertheless, full control of nonperturbative QCD effects may be hard to achieve~\cite{Ciuchini:2021smi,Gubernari:2023puw,Gubernari:2024ews}. While the claim of possible LFUV in \PB meson decays has largely disappeared in the latest LHCb publications~\cite{LHCb:2022qnv,LHCb:2022vje}, the interest in \bsll decays in general and in potential LFUV in these processes remains strong~\cite{Patnaik:2023ins}. 

In this paper we describe a search for LFUV in \BKll decays using data collected by the CMS experiment at the LHC in 2018. A special trigger and storage strategy was used to collect a large unbiased sample of $\sim$10 billion \PQb hadron decays. We also report a measurement of the branching fraction of the \BKmm decay, both differentially over the dimuon mass squared \qsq and integrated over two different \qsq ranges. The numeric values corresponding to figures and tables in this paper can be found in the \textsc{HEPdata} database~\cite{hepdata}.

This paper is organized as follows. After discussing the measurement strategy in Section~\ref{sec:strategy}, followed by the detector description in Section~\ref{sec:CMS}, we introduce the data and simulated samples in Section~\ref{sec:samples}, and event reconstruction in Section~\ref{sec:analysis}. We discuss the maximum likelihood fit to the $\PKp\ell^+\ell^-$ invariant mass distributions in Section~\ref{sec:fit}, followed by the description of the systematic uncertainties in Section~\ref{sec:syst}. The results are presented in Section~\ref{sec:results}, followed by the summary of the paper in Section~\ref{sec:summary}. Additional plots and details of the \RK measurement formalism can be found in Appendix~\ref{sec:appendix}.

\section{Measurement strategy \label{sec:strategy}}
To maximize the sensitivity to LFUV in \bsll decays, it is advantageous to use an observable that minimizes the theoretical uncertainty in its prediction. Given that the absolute \bsll rates are poorly understood because of the potentially large long-range corrections~\cite{Ciuchini:2021smi}, a more robust variable is the ratio of muon to electron decays within a certain range of the dilepton mass squared, $\qsqmin < \qsq < \qsqmax$:
\begin{linenomath}
\begin{equation}
\label{eq:RKp}
    \RK_{\text{theory}}[\qsqmin,\qsqmax]=\frac{\BR(\BKmm)[\qsqmin,\qsqmax]}{\BR(\BKee)[\qsqmin,\qsqmax]}.
\end{equation}
\end{linenomath}
This ratio is very close to unity in the SM~\cite{Hiller:2003js,Bordone:2016gaq,Isidori:2020acz,Isidori:2022bzw} and known to a precision of about 1\%. Furthermore, the experimental systematic uncertainties related to signal reconstruction and selection can be reduced by measuring \RK as a double ratio normalized to the
corresponding \BKJp decay channels:
\begin{linenomath}
\ifthenelse{\boolean{cms@external}}
{ 
  \begin{equation}
  \label{eq:RKJ}
    \RK(\qsq)[\qsqmin,\qsqmax]=[\frac{\left[\frac{\BR(\BKmm)[\qsqmin,\qsqmax]}{\BR(\BKJpmm)}\right]}
    {\left[\frac{\BR(\BKee)[\qsqmin,\qsqmax]}{\BR(\BKJpee)}\right]} .
  \end{equation}
} 
{ 
  \begin{equation}
  \label{eq:RKJ}
    \RK(\qsq)[\qsqmin,\qsqmax]=\left.\frac{\BR(\BKmm)[\qsqmin,\qsqmax]}{\BR(\BKJpmm)} \middle/
    \frac{\BR(\BKee)[\qsqmin,\qsqmax]}{\BR(\BKJpee)} \right. .
\end{equation}
}  
\end{linenomath}

This definition of \RK benefits from the fact that the branching fractions $\BR(\JPsi \!\to\! \MM)$ and $\BR(\JPsi \!\to\! \EE)$ are measured to be the same within a precision of about 0.7\%~\cite{Workman:2022ynf}, which ensures that this extra normalization does not change the value of \RK with respect to the definition in Eq. (\ref{eq:RKp}). Additionally, since the leptons from the nonresonant \BKll and the resonant \BKJpll decays occupy a similar phase space, with a similar acceptance and efficiency after the final selection, most of the systematic uncertainties related to the lepton momentum scale and identification cancel in Eq.~(\ref{eq:RKJ}), making it an excellent experimental observable directly related to $\RK_{\text{theory}}$.

The \RK ratio in Eq.~(\ref{eq:RKJ}) is measured in the \qsq region from 1.1 to 6.0$\GeV^2$, referred to as the ``low-$\qsq$'' region, which is free of dilepton resonances. This choice is optimized to exclude the contributions of the $\PGfP{1020}$ and lighter resonances at the lower end of the \qsq spectrum, and the tail of the \JPsi resonance at the higher end. In addition, the chosen lower boundary reduces the kinematic difference between the muon and electron decay channels due to the differences in their masses.  

For the differential branching fraction measurement in the \BKmm channel, the \qsq range is extended downward by adding a $\qsq < 0.98\GeV^2$ bin, below the $\PGfP{1020}$ resonance, and upward by introducing nine bins up to the kinematic limit of $\qsq = 22.9\GeV^2$, excluding the $\JPsi$ and $\PGyP{2S}$ resonance \qsq ranges. Furthermore, the low-$\qsq$ region is split into five finer bins. The definition of the 15 \qsq bins used for the differential measurement is given in Section~\ref{sec:results} (Table~\ref{tab:dif_yield}). The chosen binning ensures a sufficient number of events per bin to reliably extract the branching fraction of the \BKmm decay and is similar to that used in the LHCb analysis~\cite{LHCb:2014cxe}, which presently dominates the precision of these differential branching fraction measurements~\cite{Workman:2022ynf}. 

In addition to the low-\qsq region, part of which is defined as the signal region (SR) as described in Section~\ref{sec:analysis}, two control regions (CRs) are defined based on the dilepton \qsq, as follows. 
\begin{itemize}
	\item \JPsi CR: the $8.41 < \qsq < 10.24\GeV^2$ range is used for the \BKJpll decay, which is the main normalization channel used in the \RK measurement.
	\item $\PGyP{2S}$ CR: the $12.6 < \qsq < 14.44\GeV^2$ range is used for the \BKPsill decay, which is the secondary normalization channel, used as an additional cross-check, with the \Rpsi ratio defined as: 
\begin{linenomath}
\ifthenelse{\boolean{cms@external}}
{ 
\begin{equation}
\label{eq:Rpsi2S}
    \Rpsi= \frac{\left[\frac{\BR(\BKPsimm)}
           {\BR(\BKJpmm)}\right]}
           {\left[\frac{\BR(\BKPsiee)}
           {\BR(\BKJpee)}\right]}.
\end{equation}
} 
{ 
\begin{equation}
\label{eq:Rpsi2S}
    \Rpsi= \left.\frac{\BR(\BKPsimm)}
           {\BR(\BKJpmm)} \middle/
           \frac{\BR(\BKPsiee)}
           {\BR(\BKJpee)}\right..
\end{equation}
}
\end{linenomath}
This ratio is very close to unity in the SM and has been measured to a precision of about 8\%, with the uncertainty dominated by the current precision of the measurement of the $\PGyP{2S} \!\to\! \MM$ branching fraction~\cite{Workman:2022ynf}.
\end{itemize}

The analysis is conducted using the ``data blinding'' concept~\cite{Roodman:2003rw}. In the muon channel, a random scale factor is initially applied to the nonresonant signal yields in each \qsq bin. In the electron channel, the $\PKp\EE$ invariant mass spectrum in the vicinity of the \PBp peak in the low-\qsq region is initially kept blind. The entire analysis is optimized using simulated samples and resonant CRs, and the unblinding of the data in both channels is done only as the final step.

\section{The CMS detector\label{sec:CMS}}
The central feature of the CMS apparatus is a superconducting solenoid of 6\unit{m} internal diameter, providing a magnetic field of 3.8\unit{T}. Within the solenoid volume are a silicon pixel and strip tracker, a lead tungstate crystal electromagnetic calorimeter (ECAL), and a brass and scintillator hadron calorimeter, each composed of a barrel and two endcap sections. Forward calorimeters extend the pseudorapidity ($\eta$) coverage provided by the barrel and endcap detectors. The ECAL consists of 75\,848 lead tungstate crystals, which provide coverage in a barrel region, $\abs{\eta} < 1.48$, and two endcap regions, $1.48 < \abs{\eta} < 3.0$.  Muons are measured in gas-ionization detectors embedded in the steel flux-return yoke outside the solenoid. The procedure followed for aligning the detector is described in Ref.~\cite{CMS:2021ime}. 

The silicon tracker measures charged particles within the range $\abs{\eta} < 3.0$. During the 2018 LHC running period, when the data used in this paper were recorded, the silicon tracker consisted of 1856 silicon pixel~\cite{Phase1Pixel} and 15\,148 silicon strip detector modules. For nonisolated particles with transverse momentum in the $1 < \pt < 10\GeV$ range, the track resolutions are typically 1.5\% in \pt and 20--75\mum in the transverse impact parameter~\cite{DP-2020-049}. Muons are measured in the $\abs{\eta} < 2.4$ range, with detection planes made using three technologies: drift tubes, cathode strip chambers, and resistive-plate chambers. Matching muons to tracks measured in the silicon tracker results in a relative transverse momentum resolution, for muons with \pt up to 100\GeV, of 1\% in the barrel and 3\% in the endcaps~\cite{CMS:2018rym}. The efficiency to reconstruct and identify muons is greater than 96\%. The electron momentum is estimated by combining the energy measurement in the ECAL with the momentum measurement in the tracker. The momentum resolution is typically better than 5\% for electrons in the range $1 < \pt < 10\GeV$. It is generally better in the barrel region than in the endcaps, and also depends on the bremsstrahlung energy emitted by the electron as it traverses the material in front of the ECAL~\cite{CMS:2020uim,CMS:2024zhe,CMS-DP-2020-021}.

Events of interest are selected using a two-tiered trigger system. The first level (L1), composed of custom hardware processors, uses information from the calorimeters and muon detectors to select events at a rate of around 100\unit{kHz} within a fixed latency of 4\mus~\cite{CMS:2020cmk}. The second level, known as the high-level trigger (HLT), consists of a farm of processors running a version of the full event reconstruction software optimized for fast processing, and further reduces the event rate before data storage~\cite{CMS:2016ngn}. 

A more detailed description of the CMS detector, together with a definition of the coordinate system used and the relevant kinematic variables, can be found in Ref.~\cite{CMS:2008xjf}.

\section{Data and simulated samples}
\label{sec:samples}
\subsection{\texorpdfstring{\PB}{B} parking data sample} 
For the purpose of this analysis, we deployed a novel trigger and data processing strategy in CMS during the 2018 data-taking period at the proton-proton $(\Pp\Pp)$ center-of-mass energy of 13\TeV, referred to as ``\PB parking''. This strategy enabled the collection of order $10^{10}$ unbiased \PQb hadron decays~\cite{CMS:2024zhe} by exploiting the fact that \PQb hadrons are predominately produced in pairs, and therefore one can trigger on one \PQb hadron of the pair using a specific decay mode (``tag-side'' \PQb hadron), while the other \PQb hadron (``probe-side'' \PQb hadron) decay is \textit{unbiased} by the trigger.

The trigger strategy relies on tag-side final states that include a muon with a relatively high \pt and a significant displacement from the $\Pp\Pp$ collision point. This strategy is motivated by the significant fraction (nearly 40\%) of \bbbar events that include at least one muon from a \PQb hadron (or subsequent charm hadron) decay, combined with the relatively low rate of L1 single-muon triggers. In addition, the \PB parking strategy takes advantage of the gradual decrease of the L1 trigger rate and online computing resources use by the nominal CMS physics program as the instantaneous luminosity \lumi decreases during each LHC fill.  To exploit the available L1 (HLT) bandwidth of up to 30 (5.4) kHz, the \PB parking trigger requirements were gradually relaxed during each LHC fill. This strategy does not affect the main, high-\pt physics program of CMS, while offering a significant increase in the CMS potential in flavor physics, including the present \RK measurement.

\begin{table*}[!tbh]
    \centering
    \topcaption{Summary of the loosest muon trigger requirements imposed by the L1 and HLT algorithms for each instantaneous luminosity scenario: the L1 and HLT muon transverse momentum thresholds \ptm, and the HLT muon impact parameter significance \ipsig.  Also shown are the trigger purity, peak HLT rate, and $\intlum$. The second trigger was the highest threshold one during early data taking, corresponding to $\intlum = 6.9\fbinv$, and then the second-highest for the rest of the data taking, accumulating $\intlum = 26.7\fbinv$ out of $34.7\fbinv$ collected by the highest threshold trigger.}
\vspace{\cmsTabSkip}
    \begin{tabular}{ccccccc}
\hline\hline
 \mystrut \lumi	 & L1 \ptm 		& HLT \ptm 	& HLT \PGm \ipsig 	& Purity 	& Peak HLT  & \intlum \\
    \lumunit       	& thr. [$\GeVns$] 	& thr. [$\GeVns$] 	& thr. 			& [\%] 	& rate [kHz]  & [$\!\fbinv$]\\
    \hline
1.7 & 12 & 12 & 6 & 92 & 1.5 & 34.7 \\
1.5 & 10 & 9  & 6 & 87 & 2.8 & $6.9 + 26.7$ \\
1.3 & 9  & 9  & 5 & 86 & 3.0 & 20.9 \\
1.1 & 8  & 8  & 5 & 83 & 3.7 & 8.3\\
0.9 & 7  & 7  & 4 & 59 & 5.4 & 6.9 \\
    \hline\hline
    \end{tabular}
    \label{tab:trigger}
\end{table*}

Table~\ref{tab:trigger} summarizes the \PB parking muon trigger requirements imposed by the L1 and HLT algorithms. The L1 trigger logic requires the presence of a muon with $\abs{\eta} < 1.5$ and with a variable minimum \pt threshold. These requirements help to control the L1 rate by removing events with muons at low \pt and large \abs{\eta}, which are dominated by those produced in pileup interactions (additional minimum bias interactions within the same or adjacent bunch crossings). In addition, a single-muon L1 trigger with $\pt > 22\GeV$ in the full pseudorapidity range $\abs{\eta} < 2.4$ is used. Since the muon displacement information is not available at L1, only \pt and $\abs{\eta}$ thresholds are used to control the rate. At the HLT, the \pt threshold is sharpened, given the more precise momentum reconstruction compared to that at L1, and a minimum requirement is imposed on the two-dimensional (2D) muon track impact parameter significance \ipsig, defined as the distance of closest approach of the track to the beam line (the measured line of the proton beams inside the CMS detector~\cite{CMS:2014pgm}), divided by its uncertainty. These triggers are ``nested'', \ie, the most restrictive trigger is exposed to the largest integrated luminosity ($\intlum$), while the looser ones have progressively higher efficiency but lower $\intlum$.

The purity of the \PB parking sample is defined as the fraction of triggered events containing a \PQb hadron decay
and is evaluated by counting the number of $\PBz \!\to\! \PDstp\PGmm\PAGnGm \!\to\! \PDz\PGp^+_{\text{soft}}\PGmm\PAGnGm \!\to\! \PKm\PGpp\PGp^+_{\text{soft}}\PGmm\PAGnGm$ events, using the known \PBz meson production fraction $f_\PQd$ and the relevant branching fractions~\cite{Workman:2022ynf}. For the $\PGp^+_{\text{soft}}$, the \pt threshold is set at 0.5\GeV. The purity is measured both in data and in simulation (described in Section~\ref{sec:MC}), with good agreement between the two estimates. The purity ranges between 59 and 92\%, depending on the trigger thresholds, with an average of $\approx$80\% in the collected data set.

The \PB parking data set corresponds to $\intlum = 41.6 \pm 1.0$\fbinv~\cite{LUM-18-001}, about 30\% less than the \intlum of 59.8\fbinv collected with the main physics triggers operating during the entire 2018 data taking. It took about a year to fully reconstruct these data during the LHC Long Shutdown 2; hence the name: \PB parking.
 
\subsection{Event simulation}
\label{sec:MC} 
Monte Carlo (MC) simulated samples are used to optimize the analysis and model various background sources. Signal and background processes are generated with \PYTHIA 8.230~\cite{Sjostrand:2014zea}, including parton showering, fragmentation, and hadronization. The \PYTHIA output is interfaced with  \EVTGEN 1.3.0~\cite{Lange:2001uf}, which simulates various \PQb hadron decays. The underlying event is also modeled with \PYTHIA, using the CP5 tune~\cite{CMS:2019csb}. The parton distribution functions (PDFs) are taken from the NNPDF3.1 set~\cite{NNPDF:2017mvq}. Final-state photon radiation is modeled with \PHOTOS 3.61~\cite{Barberio:1993qi}. The \BKll signal is generated with the BTOSLLBALL, setting 6, model of the \EVTGEN decay  library, based on form-factors from Ref.~\cite{Ball:2004rg}. While more precise form-factor determinations are now available both from light-cone sum rules~\cite{Bharucha:2015bzk,Gubernari:2018wyi} and lattice QCD~\cite{Parrott:2022rgu}, we do not expect them to have a significant impact on the decay kinematics. Additionally, the following processes are simulated: the \BKJp and \BKPsi resonant decays, and \BKstzJp, $\PBz \!\to\! \PGyP{2S}\PKstz$, $\PBz \!\to\! \PKstz\ell^+\ell^-$, \BKstpmJp, $\PBp \!\to\! \PGyP{2S}\PKstp$, and $\PBp \!\to\! \PKstp\ell^+\ell^-$ backgrounds, as well as inclusive \PQb hadron samples, and Cabibbo-suppressed $\PBp \!\to\! \JPsi \PGpp$, $\PBp \!\to\! \PGyP{2S}\PGpp$, and $\PBp \!\to\! \PGpp\ell^+\ell^-$ decays, with and without emulation of the trigger requirements. The CMS detector response is simulated using \GEANTfour~\cite{Agostinelli:2002hh}. All MC samples are reconstructed with the same software packages as used for collision data and include effects of pileup by overlaying simulated minimum bias events on the hard-scattering event, with the multiplicity distribution matching that in data.

\section{Event reconstruction and selection\label{sec:analysis}}
The particle-flow (PF) algorithm~\cite{CMS:2017yfk} aims to reconstruct and identify each individual particle (muon, electron, photon, and charged and neutral hadrons) in an event, with an optimized combination of information from the various elements of the CMS detector.

Muons are identified as tracks in the silicon tracker consistent with either a track or several hits in the muon system, and associated with calorimeter deposits compatible with the muon hypothesis~\cite{CMS:2018rym}. A muon that triggered the event readout is required to pass the ``medium'' muon identification criteria of Ref.~\cite{CMS:2018rym}. As multiple $\Pp\Pp$ collision vertices are reconstructed from each beam crossing, we select a single primary vertex (PV) as the one whose $z$ position is closest to that of the point of closest approach of this muon to the beam line. In addition, for the electron channel, only electron candidates consistent with originating from this PV are considered. This procedure significantly reduces the contamination from pileup interactions. To simplify the trigger efficiency calculations, the muon channel analysis uses only one HLT path, which requires a muon with $\pt > 9\GeV$ and $\ipsig > 6$. This trigger was the highest threshold one during early \PB parking data taking, corresponding to $\intlum = 6.9\fbinv$, and then the second-highest for the rest of the data taking, accumulating an additional $\intlum = 26.7\fbinv$. Therefore, the \intlum used for the muon channel is $33.6\fbinv$, as shown in the second row of Table~\ref{tab:trigger}.

The PF algorithm used for electron reconstruction imposes an implicit minimum electron $\pt$ requirement of $2\GeV$. To recover the efficiency loss for low-\pt (LP) electrons, an additional algorithm (``LP electron reconstruction'') was specifically developed. As is the case for the PF algorithm, in the LP algorithm, the determination of the charged-particle track parameters for electron candidates, in the presence of bremsstrahlung energy loss, relies on the use of a Gaussian sum filter~\cite{CMS:2020uim}. The LP electron candidate reconstruction is started with a combination of two boosted decision trees (BDTs), which are trained on samples of low-$\pt$ electrons and result in looser reconstruction requirements than for the PF electrons. The LP electrons in this analysis are reconstructed down to 1\GeV, which ensures that they reach the ECAL\@. In case of duplicates, we preferentially select PF electrons over LP ones.  To optimize the performance of the electron identification, several electron candidate characteristics are combined using a BDT into a single discriminating variable, referred to as ID, analogously to what was done in Ref.~\cite{CMS:2020uim}. The ID BDTs are trained separately for PF and LP electrons. The PF electron ID BDTs were retrained specifically for this analysis to improve the performance at low \pt. The training was done separately for two ranges, $2 < \pt < 5\GeV$ and $\pt > 5\GeV$, while a single training is used for the LP electron ID in the entire \pt range. The input variables for the ID BDTs include both track-related quantities and calorimetric shower shapes, as well as variables related to the matching of the extrapolated track to the calorimeter cluster and the difference between the track momentum at the innermost and outermost tracker layers. The electron ID BDTs were trained with the \textsc{XGBoost} algorithm~\cite{xgboost} on a simulated sample of \BKJpee events. The electron \pt and $\eta$ distributions of the training sample have been reweighted to reproduce those for the background, to avoid biases. The ``tag-and-probe'' method~\cite{CMS:2010svw} using $\JPsi \!\to\! \EE$ decays in data is used to check the accuracy of the simulation for the ID BDT input variables, as well as for the output distribution. We find these variables for both the PF and LP electrons in data to be consistent with those in simulation, within statistical uncertainties. 

To measure \RK, we select events for which either a \BKmm candidate (used in the numerator of the \RK ratio) is found on the tag side, \ie, a muon from the \BKmm candidate must satisfy the trigger conditions, or a \BKee candidate (used in the denominator of the \RK ratio) is found on the probe side. While using the tag side for the muon channel and probe side for the electron channel complicates the analysis, as the trigger efficiency does not cancel in the \RK ratio, this choice ensures a large number of events collected in the \BKmm channel and not only maximizes the sensitivity of the \RK measurement, but also enables high-precision measurements of the \BKmm branching fraction. 

The \PBp candidates are formed using a pair of opposite-sign (OS) same-flavor leptons with an invariant mass below $5\GeV$ and a positively charged track, to which the kaon mass is assigned. Several quality criteria are applied to each muon, electron, and track candidates to reduce the number of misreconstructed objects. Specifically, tracks are required to pass the ``high-purity'' track quality criteria~\cite{CMS:2010vmp,CMS:2017yfk} with $\pt > 1\GeV$ and $\abs{\eta} < 2.4$. Since the triggering object is a muon, to reconstruct the \BKmm decay only one additional muon is needed. This muon is required to have OS with respect to the triggering muon, pass the ``medium'' muon identification criteria, have $\pt > 2\GeV$, and satisfy $\abs{\eta} < 2.4$. In the electron channel, in order to increase the purity of the sample, we require at least one of the electrons to be reconstructed with the PF algorithm. Both electrons are required to have $\abs{\eta}<2.4$. The two electrons must have points of origin separated, along the beam line direction, by $\abs{\Delta z}<1\unit{cm}$ and with $\abs{\Delta z}<1\unit{cm}$ of the point of origin of the muon that triggered the event. At this stage, we apply only very loose requirements on the ID BDT output for the electrons, as it is used as an input to the multivariate analysis described below. The transverse momentum of the \PBp candidate is required to exceed 3 and 1.75\GeV in the muon and electron channels, respectively.

The tracks of the three particles forming the \PBp candidate are fitted to a common vertex, using their measured momentum vectors with the corresponding uncertainties, to improve the mass measurement accuracy for both the \PBp candidate and the lepton pair.  The kinematic fit algorithm~\cite{kinFit} constrains the tracks belonging to the \PBp candidates to originate from a single vertex, and provides the vertex position, covariance matrix, and $\chi^2$ of the fit. After the vertexing is performed, the particle trajectories are refitted using this secondary vertex (SV) as an additional constraint, and their momenta are recomputed. The masses of the particles (leptons and kaon) are fixed to their nominal values~\cite{Workman:2022ynf}. A loose set of criteria is imposed on the SV fit probability, as well as the $L_{xy}/\sigma_{xy}$ and \cosa variables described in Table~\ref{tab:bdt}. The invariant mass of the \PBp candidate is required to be in the range 5.0--5.6 (4.7--5.7)\GeV for the muon (electron) channel.  In addition, in the electron channel, the two electron candidates are fitted to a common vertex (``dielectron vertex''), which is used later in the analysis to define one of the discriminating variables, $\dthreeD$, described in Table~\ref{tab:bdt}.

A significant fraction of semileptonic decays of heavy-flavor hadrons, containing a semileptonic charm meson decay or a hadronic $\PDz \!\to\! \PKm\PGpp$ decay, remains in the preselected sample. The hadronic \PDz decays enter the sample  because of misidentification of the \PKm or \PGpp meson as a lepton. Therefore, a charm veto is applied by requiring that, for a \PBp candidate, the invariant mass of the track, using the pion mass assignment, and the OS lepton, using the kaon mass assignment, is larger than 2\GeV. In the electron channel, we additionally require the invariant mass of the track, using the kaon mass assignment, and the OS electron, using the pion mass assignment, to also be larger than 2\GeV. This additional selection is not applied in the muon channel, as the probability to reconstruct charged pions as muons is significantly smaller than that for kaons. Finally, to suppress events where a muon is misreconstructed as a kaon candidate, the track under the muon mass hypothesis and the OS muon belonging to a \PBp candidate are required to have an invariant mass away from known dimuon resonances.

The selected \PBp candidates are binned in \qsq, as described in Section~\ref{sec:strategy}. Events in each \qsq bin can be further divided in regions based on the invariant mass of the \PBp candidate. The region where genuine \PBp candidates are expected, $5.07 < \mKll < 5.49\GeV$, is denoted as the SR\@. The adjacent mass regions $4.90 < \mKmm < 5.07$ ($4.91 < \mKee < 5.07$)\GeV and $5.49 < \mKll < 5.65\GeV$ are designated as sidebands (SBs). The low-mass SB has a significant contamination from partially reconstructed \PB meson decays, while the high-mass SB is dominated by  combinatorial background. 

The final selection in each channel is based on a BDT, which combines several variables into a classifier built using the \textsc{XGBoost} package. The input variables used in the BDT are selected based on the forward elimination method (\ie, by adding one variable at a time and either keeping or dropping it based on the observed improvement in the performance) and are summarized in Table~\ref{tab:bdt}. In the electron channel, two independent BDTs are trained for the PF-PF and PF-LP categories using the same input variables. Given the significantly higher background in the electron channel, the corresponding BDTs have more input variables and have all the momenta variables normalized to the \PBp candidate mass \mKee to ensure that a BDT score selection does not introduce peaks in the invariant mass distributions (``mass sculpting''). Both muon and electron BDTs are tested thoroughly using data and simulated events to ensure that they do not introduce any mass sculpting. The BDTs are trained in a supervised manner with simulated \BKll decays in the low-\qsq bin as signal and data events from the low-\qsq bin SBs as background. 

\begin{table*}[!htb]
	\centering
	\topcaption{Input variables used in the muon and electron channel BDTs.}
	\begin{tabular}{m{0.22\textwidth}m{0.73\textwidth}}
		\hline\hline
		Variable & Description \\
		\hline
		\multicolumn{2}{c}{Common variables}\\
		\hline
  	 $\cosa$ & Cosine of the angle in the plane transverse to the beams between the momentum vector of the \PBp candidate and the line connecting the beam line and the SV.\\[3\cmsTabSkip]
  	 $p(\text{\PBp~vtx})$ & Probability of the SV kinematic fit.\\[0.5\cmsTabSkip]
  	 $L_{xy}/\sigma_{xy}$ & Significance of the SV displacement in the transverse plane with respect to the beam line.\\[\cmsTabSkip]
  	$\pt(\PBp)$& Transverse momentum of the \PBp candidate; in the electron channel it is divided by $\mKee$.\\[\cmsTabSkip]
  	$\pt(\PKp)$& Transverse momentum of the \PKp candidate;  in the electron channel it is divided by $\mKee$. \\[2\cmsTabSkip]
		\hline
		\multicolumn{2}{c}{Muon channel variables}\\
		\hline
  	$\min \Delta R (\PGm, \PKp)$& \mystrut$\Delta R = \sqrt{\smash[b]{(\Delta\eta)^2 + (\Delta\phi)^2}}$ distance between the \PKp candidate and the closest muon candidate.\\[2\cmsTabSkip]
  	$\min \Delta z (\PGm, \PKp)$ & $\Delta z$ distance between the points of origin of the \PKp candidate and the closest muon candidate along the beam line direction.\\[2\cmsTabSkip]
  	Iso$(\mu_{\text{lead}})$ & PF isolation for the \pt-leading muon candidate, defined as a scalar \pt sum all PF candidates, excluding the muon candidate itself, within $\Delta R< 0.4$ of the muon candidate and corrected for pileup.\\[\cmsTabSkip]
		\hline
		\multicolumn{2}{c}{Electron channel variables}\\
		\hline
         	\mystrut$\pt(\Pe_{i})/\mKee,\newline i = 1, 2$ & Transverse momenta of the two electron candidates, divided by $\mKee$. \\[2\cmsTabSkip]
         $\Delta z(\Pe_{i}, \PKp), \;i = 1, 2$ & Longitudinal distance between the points of origin of each electron candidate and the kaon candidate. \\[\cmsTabSkip]
         $\smash[b]{\frac{\dthreeD}{\sigma_{\dthreeD}}}$ & Kaon candidate 3D impact parameter significance with respect to the dielectron vertex. \\[2\cmsTabSkip]
          $\Delta R(\Pep, \Pem)$ &  $\Delta R$ between the two electron candidates.  \\[0.5\cmsTabSkip]
         $\Delta R(\Pe_{i}, \PKp), \;i = 1, 2$ &  $\Delta R$ between each electron candidate and the kaon candidate.\\[0.5\cmsTabSkip]
         $\frac{|\mathbf{p(\EE)}\times \mathbf{r}| - |\mathbf{p(\PKp)}\times \mathbf{r}|}{|\mathbf{p(\EE)}\times \mathbf{r}| + |\mathbf{p(\PKp)}\times \mathbf{r}|}$ &  Asymmetry of the momentum of the dielectron system and that of the $\PKp$ momentum  with respect to the \PBp candidate trajectory, where $ \mathbf{r}$ is a unit vector connecting the PV and SV. \\[3\cmsTabSkip]
         ID($\Pe_{i}, \;i = 1, 2$) &  Electron ID BDT score for two electron candidates.  \\[0.5\cmsTabSkip]
         $I_{\Delta R=0.4}^{\text{rel}}(\Pe_i), \; i = 1,2$\newline\mystrut and $I_{\Delta R=0.4}^{\text{rel}}(\PKp)$ &  Relative track-based isolation of the two electron candidates and the $\PKp$ candidate, respectively, defined as a scalar \pt sum of all additional tracks in a $\Delta R<0.4$ cone around the candidate, divided by the candidate's $\pt$. \\[2\cmsTabSkip]
  	\hline\hline
\end{tabular}
\label{tab:bdt}
\end{table*}

\begin{figure*}[!htb]
	\centering
	\includegraphics[width=0.327\textwidth]{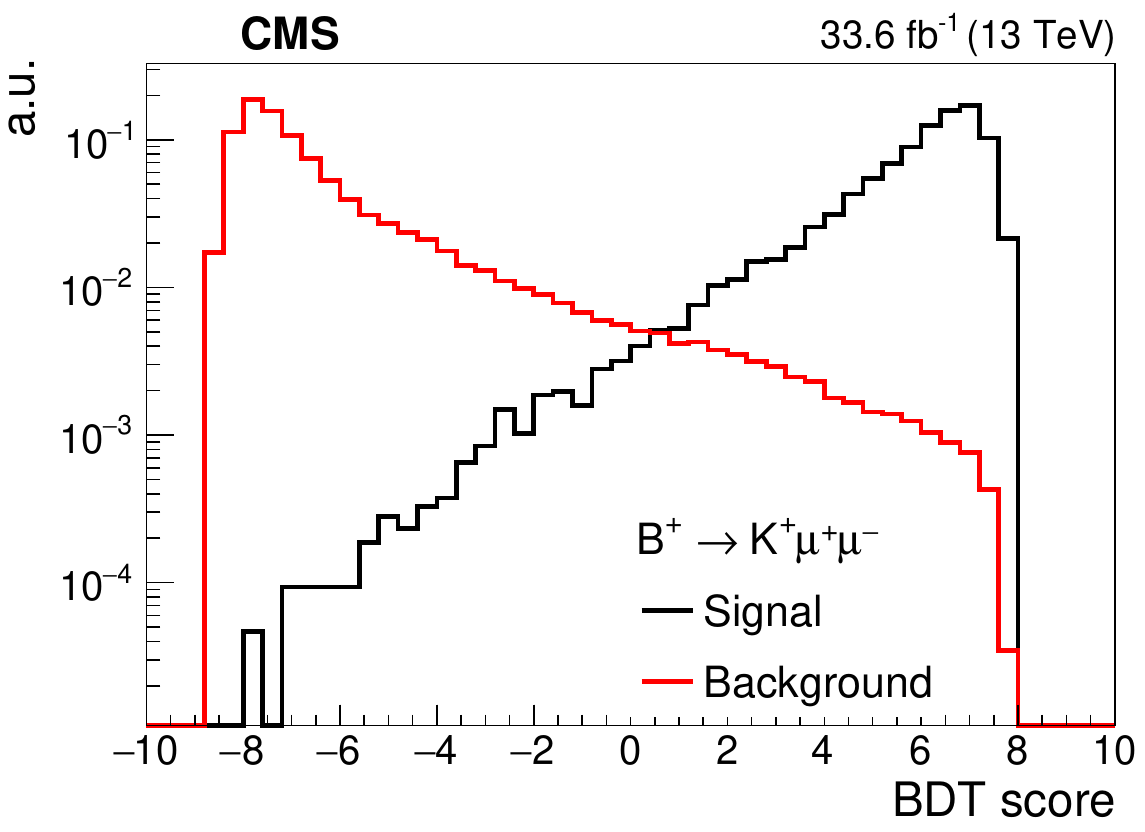} 
	\includegraphics[width=0.327\textwidth]{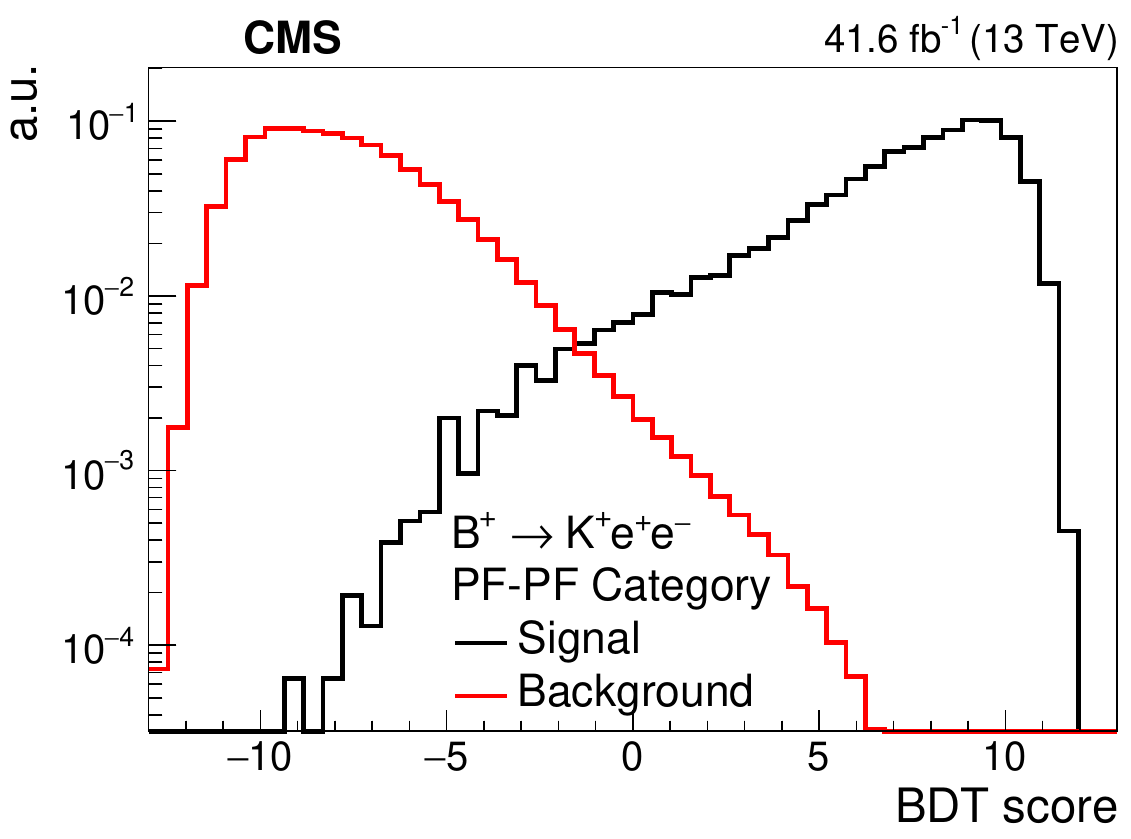} 
	\includegraphics[width=0.327\textwidth]{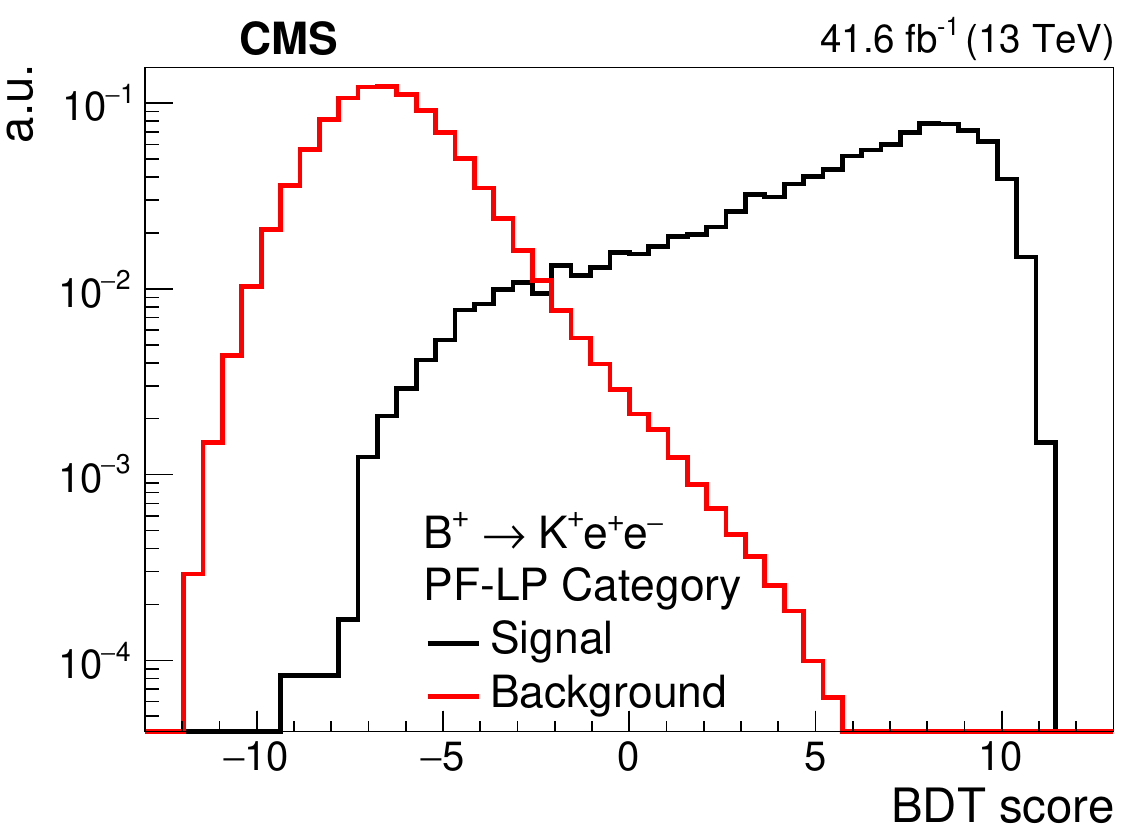} 
	\caption{Analysis BDT output for signal (MC simulation, in black) and background (same-sign dilepton data, in red) for the muon channel (left) and for the PF-PF (center) and PF-LP (right) electron channels. The histograms are normalized to unit area.}
	\label{fig:bdt}
\end{figure*}

It is checked that the simulation describes the data well. The comparisons are performed for all BDT input variables, as well as the BDT output, using the \splot technique~\cite{Pivk:2004ty}, where the reconstructed \PBp candidate invariant mass is used as the discriminating variable. The test is performed on the \BKJpll CRs, and good agreement between data and simulation is found for all three channels (muon, PF-PF, and PF-LP). After training, the BDT output is a continuous function, as shown in Fig.~\ref{fig:bdt}, where high (low) values correspond to signal-like (background-like) events. In this figure, the background sample is obtained by using the SR selection, with the inversion of the OS requirement for the two leptons (the same-sign CR). The optimal BDT working point (WP) is chosen to maximize the expected significance of the \BKll signal in the low-\qsq region. In the muon channel, the same BDT selection is applied to the $\JPsi$ and $\PGyP{2S}$ CRs to allow for maximum cancellation of the systematic uncertainties, while in the electron channel, where the dominant uncertainty is statistical, a looser BDT WP is used for the resonance CRs to maximize the number of events in the normalization channel. The muon, PF-PF, and PF-LP channels are then combined statistically for the \RK extraction.

The product of the detector acceptance $\mathcal{A}$ and efficiency $\epsilon$ is evaluated using simulated samples without the trigger requirements. For the muon channel, the trigger efficiency \etrig, determined from simulated samples with and without the trigger requirement and no reconstruction requirements, is found to be approximately 0.4\%. The low value of the trigger efficiency is mainly due to the muon \pt and displacement requirements. In the electron channel, since the \BKee candidates are found on the probe side, the \etrig component does not enter in the \RK extraction, as it is measured to be the same for the low-\qsq and the \JPsi regions and hence cancels completely in the \RK ratio.

Slight differences between data and simulation in the muon channel are mitigated with scale factors (SFs) applied to simulated events, and the correction for the following effects are applied.
\begin{itemize}
	\item Trigger efficiency: The difference in the trigger response is corrected for by an SF, as a function of muon $\eta$, \pt, and the SV displacement significance $L_{xy}/\sigma_{xy}$. The SF is measured with a tag-and-probe method~\cite{CMS:2010svw}, exploiting the $\JPsi \!\to\! \MM$ decays. 
	\item Muon identification efficiency: The quality criteria imposed on the muons may have different performance in data and simulation. This SF is measured with the tag-and-probe method.
	\item BDT efficiency: The BDT is using input variables that have small differences between data and simulation. The corresponding SF is measured by comparing the BDT output for the \BKJpmm events in data and simulation.
	\item Pileup: Simulated events are reweighted to match the evolving pileup conditions of the analyzed data.
	\item Higher-order corrections: Leading-order \PYTHIA simulation is reweighted at the generator-level so that the \PB meson \pt and rapidity spectra match those from fixed-order perturbative quantum chromodynamics calculations with resummation of large logarithms that arise from soft and collinear emissions~\cite{Cacciari:1998it,Cacciari:2001td}. The reweighting is done using the fixed-order plus next-to-leading logarithmic (FONLL) calculations~\cite{Cacciari:2012ny,Cacciari:2015fta}, which are known to reproduce these spectra in data with high accuracy.
\end{itemize}
It was found that applying the analogous SFs to the electron channel changes the measured value of \RK by less than 1.2\%, which has a negligible effect on the final result given the large statistical uncertainty in the electron channel. Therefore, the SFs are not applied in the electron channel. The measured $\Aepst$ in the muon channel for all \qsq bins, after applying these corrections, is shown in Fig.~\ref{fig:acceff_muon}. The \Aepst or \Aeps values in the low-\qsq regions and two resonant CRs for the muon or electron channel are listed in Table~\ref{tab:final_ae} .

\begin{table*}[!htb]
	\centering
	\topcaption{The product of acceptance, and offline and trigger efficiency (\Aepst) for the signal in the low-\qsq region and for the two resonance CRs. In the case of electrons, the trigger efficiency is not included in the quoted \Aeps numbers, as it cancels out in the \RK double ratio. Uncertainties are statistical only.}
	\begin{tabular}{lccc}
	\hline\hline
		\qsq range\mystrut & Muon \Aepst [\%] & PF-PF \Aeps [\%] & PF-LP \Aeps [\%] \\
		\hline
		Low \qsq\mystrut & $0.0825 \pm 0.0013$ & $2.95 \pm 0.07$  & $1.50 \pm 0.05$\\  
		\JPsi  & $0.0969 \pm 0.0006$ & $2.48 \pm 0.03$ & $1.08 \pm 0.01$\\
                $\PGyP{2S}$  & $0.1112 \pm 0.0004$ & $1.82 \pm 0.02$ & $0.68 \pm 0.01$\\
		\hline\hline
	\end{tabular}
	\label{tab:final_ae}
\end{table*}

\section{Mass fit\label{sec:fit}}
In each channel, the \BKll signal yield is extracted from an unbinned maximum likelihood fit to the invariant mass spectrum of all \PBp candidates that pass the selection criteria. Signal and background shapes are described by analytical functions or templates based on studies of simulated events, with their normalization factors free to vary in the fit to data. To allow a better description of the data, some parameters of the analytical functions are constrained within the uncertainties obtained from a fit to simulated samples, rather than being fixed. The same functional forms describing the various contributions are shown to work well in all \qsq bins in simulation; hence they are kept the same in the fit to data as well.

\subsection{Background composition}
While the background composition differs from one \qsq bin to another and from channel to channel, the background originates from the following common sources: 
\begin{itemize}
	\item Combinatorial background: This background stems from the combination of objects from different \PQb hadron decays, or two muons from the same \PQb hadron decay chain, with a random track. The shape is extracted by exploiting the fact that this background dominates the upper \PBp mass SB, as well as the same-sign CR.
	\item Partially reconstructed background: This background is dominated by the $\anyB \!\to\! \Kstar\PX$ decays, where \PX is \JPsi, $\PGyP{2S}$, or a nonresonant $\ell^+\ell^-$ pair. This background arises from $\PKstz \!\to\! \PKp\PGpm$, $\PKstp \!\to\! \PKp\PGpz$, and $\PKstp \!\to\! \PKz \PGpp$ decays, where either the \PKp or \PGpp track is used together with the dilepton system to build a \PBp candidate, with the other decay particle being either lost or ignored. In the latter case, potentially two \PBp candidates are reconstructed by combining the dilepton system with either of the two tracks; both have significantly lower mass than the nominal \PBp meson mass due to the missing other particle, but still potentially contaminate the signal window in the $\PK\ell^+\ell^-$ mass. Simulated $\anyB \!\to\! \Kstar\PX$ events are used to derive the invariant mass distributions of the partially reconstructed backgrounds.
	\item Cabibbo-suppressed background: This background is present in both the resonant CRs and nonresonant SR, and it is very similar to the signal signature, just with the kaon replaced by a pion. This background cannot be discriminated from the signal decay and it is accounted for using simulated events.
	\item Resonant background ``leakage'': This background comes from the fact that leptons produced in the \BKJpll and \BKPsill decays can radiate final-state photons. In the case of muons, no photon recovery algorithm is used, so these photons are not accounted for in the muon momentum reconstruction; in the case of electrons, while most of these photons are picked by the Gaussian sum filter algorithm and added to the momentum of the reconstructed electrons, large-angle radiation may still be missed. These effects reduce the \qsq to lower values and generally result in migration of events across the \qsq bins. The finite detector resolution and a much larger branching fraction to the resonant mode can also contribute tails on the high side. This background source is significant only in the \qsq bins in the vicinity of the \JPsi and $\PGyP{2S}$ resonances, and is estimated using simulated samples.
	\item Other \PQb hadron decays: This is the most general background that includes all partially reconstructed \PQb hadron decays that do not fit in any of the previous categories. Depending on the \qsq bin, different specific decays dominate in this category. The shape of this background was evaluated using an inclusive sample of soft QCD processes generated with \PYTHIA, with a filter that selected \PQb hadron production. The \mKll distribution in this sample was found to be well described by a falling exponential function in all \qsq bins.
\end{itemize}

A potential peaking background in the electron channel could arise from hadrons misidentified as electrons in the all-hadronic decays of the type $\PBp \!\to\! \PKp \mathrm{h_1h_2}$, where h$_{1,2}$ are hadrons, dominated by $\PBp \!\to\! \PKp\PGpp\PGpm$ decays (with the branching fraction of ${\sim}10^{-5}$, once the resonant charm $\PADz\PGpp$ contribution is subtracted). The peaking structure is particularly pronounced in the low-\qsq region where the dipion system generally has large momentum, and replacing it with the dielectron hypothesis consequently does not shift sizably the reconstructed invariant mass. This background, missed in early \RK measurements~\cite{LHCb:2014vgu,LHCb:2019hip,LHCb:2021trn}, was largely responsible for the claimed anomaly, and after accounting for it, the \RK value is found to be compatible with the SM expectation~\cite{LHCb:2022qnv,LHCb:2022vje}. In the present analysis, we have developed the identification of low-\pt electrons and optimized the identification of PF electrons specifically to reduce the probability of such misidentification. A typical misidentification rate after the BDT selection is $10^{-3}$--$10^{-4}$ per electron, making the misidentified-hadron background very small. Nevertheless, given the importance of the peaking background in the LHCb case, we have made an explicit estimate of potential contributions of the misidentified-hadron background.  We use a large simulated sample of $\PBp \!\to\! \PADz(\!\to\! \PKp\PGpm)\PGpp$ decays without applying the charm veto (described in Section~\ref{sec:analysis}) to estimate the misidentification efficiency.  In addition, we use a small simulated sample of charmless $\PBp \!\to\! \PKp\PGpm\PGpp$ decays to evaluate the charm veto efficiency for this mode.  The study showed that the charm veto plus the identification requirements on PF and LP electrons work very well to suppress the misidentification background to a low level, as expected from squaring the misidentification probability and applying the relative branching fractions for the signal and background processes. The total estimated number of misidentified-hadron background events in the low-\qsq region after the full analysis selection is less than 0.3 events in each of the PF-PF and PF-LP categories. Compared to the expected signal yield, this peaking background corresponds to $<$2 and $<$6\%, respectively, for the two categories, which is much smaller than the corresponding statistical uncertainty. As this background is also shown to be negligible in the muon channel, it is not included in the final results. 

\subsection{Muon channel signal extraction}
In the muon channel, two independent unbinned maximum likelihood fits are performed to the \PBp candidate invariant mass distribution: a ``single'' fit in the low-\qsq region and a ``simultaneous'' fit across the 15 \qsq bins. The former is used to measure \RK and the integrated branching fraction of the \BKmm decay, while the latter is used for the extraction of the differential branching fraction. While the two fits give consistent results in the low-\qsq region, both in terms of the central value and the uncertainty, the strategy of having two fits achieves greater similarity between the \BKmm and \BKee channels and reduces the \RK measurement uncertainty. 

In both the single fit of the low-\qsq region and the simultaneous fit, the signal is described with the sum of a double-sided Crystal Ball (DCB) function~\cite{Oreglia,Gaiser} and a Gaussian function. All of the parameters of the DCB function are fixed based on simulation. The mean and the width of the Gaussian function in the single fit are free parameters. In the simultaneous fit, while the width is allowed to float, the mean of the Gaussian function is parameterized as a linear function of \qsq with a slope of $0.0014 \pm 0.0012\GeV^{-1}$ to describe the observed dependence in simulation. The rest of the parameters in the simultaneous fit are treated the same way as in the single fit.

\begin{itemize}
\item The $\PBp \!\to\! \Kstar\PX$ background is described by a DCB function with the shape fixed from simulation. In the simultaneous fit, an exponential function with parameters fixed from simulation is added to the DCB function to account for the \PBp candidates built with the \PGp of the $\PKst(892)$ decay in the 11.0--19.24$\GeV^2$ \qsq range where the contribution of this component is significant. 

\item The sum of the combinatorial and  other \PBp meson decays backgrounds is described with a single exponential function, except for the last \qsq bin of the simultaneous fit, where it is multiplied by $\mKmm - \mmm - m_\PKp$ to account for the phase space suppression. 

\item The $\PBp \!\to\! \PGpp\PX$ background is described by a DCB function with the shape and relative yield with respect to the $\PBp \!\to\! \PKp\PX$ signal fixed from simulation. 

\item The \JPsi ($\PGyP{2S}$) resonant background leakage to nearby \qsq bins is described with a DCB function with the normalization as a free parameter of the fit and the shape parameters fixed by fitting simulated \BKJpmm (\BKPsimm) events in the specific \qsq bin. This contribution is included only in the simultaneous fit in the following \qsq bins: 6.0--7.0 and 7.0--8.0$\GeV^2$ for the \BKJpmm background, and 11.0--11.8, 11.8--12.5, and 14.82--16.0$\GeV^{2}$ for the \BKPsimm background. 
\end{itemize}

For the \BKJpmm CRs, the fit is kept as close as possible to the single fit in the low-\qsq region in terms of template functions and parameter treatment. Nevertheless, because of the much smaller radiative tail due to the tight $2.9 < \mmm < 3.2\GeV$ requirement in the \JPsi CR and the different background sources, some of the templates are different from the ones used for nonresonant signal. The signal is described by a sum of three Gaussian functions with all shape parameters constrained within the uncertainties from a fit to simulated data. The \BKstJpmm background is described with the sum of a DCB function and an exponential function, with all the shape parameters fixed from simulation, as in the 11.0--19.24$\GeV^2$ \qsq range of the simultaneous fit.

The \BKPsimm CR is not used directly in the analysis, but is utilized for several cross checks (such as the \Rpsi measurement). For validation purposes, the functions used for the fit in the \BKPsimm CR are exactly the same as in the low-\qsq region, except for the \BKstJpmm background, which is described with the sum of a DCB function and an exponential function with all the shape parameters fixed from simulation, as in the 11--19.24$\GeV^2$ \qsq range of the simultaneous fit.

\begin{figure*}[!htb]
	\centering
 		\includegraphics[width=0.47\textwidth]{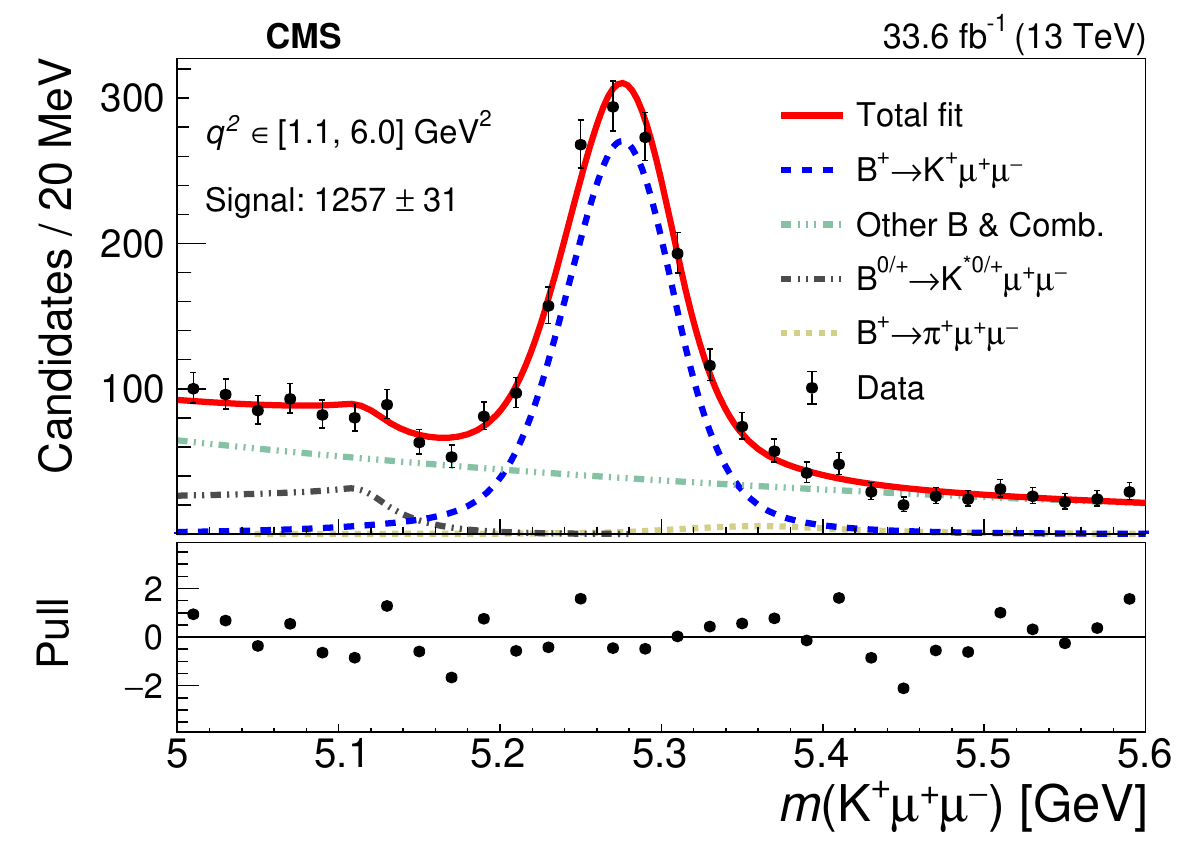}\\
		\includegraphics[width=0.47\textwidth]{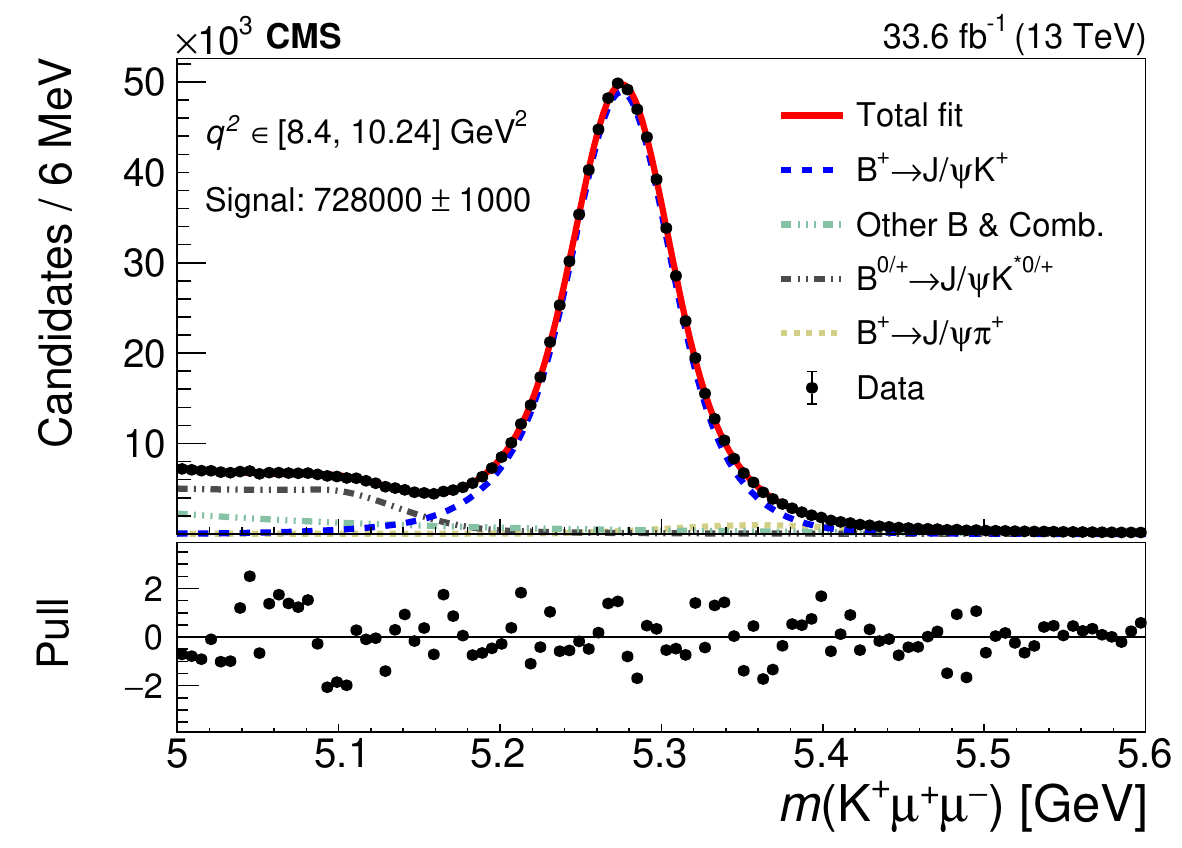}
		\includegraphics[width=0.47\textwidth]{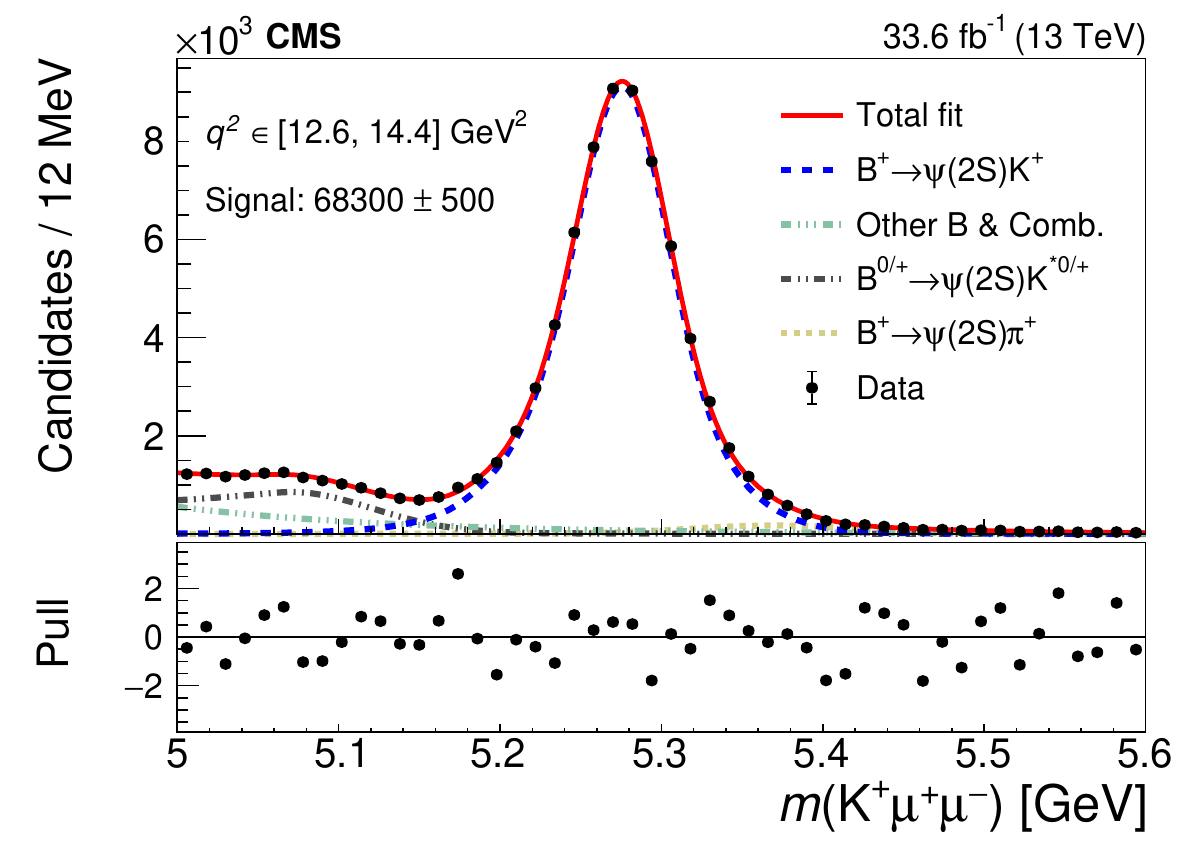} 
 	\caption{Results of an unbinned likelihood fit to the $\PKp\MM$ invariant mass distributions in the low-\qsq bin (upper), and in the \BKJpmm (lower left) and \BKPsimm (lower right) CRs. The error bars show the statistical uncertainty in data. The lower panels show the distribution of the pull, which is defined as the Poisson probability to observe the number of event counts in data, given the fit function, expressed in terms of the Gaussian significance.}
	\label{fig:fit_muons}
\end{figure*}

The functions used in the fit are summarized in Table~\ref{tab:fit_function_muons}. The results of the unbinned maximum likelihood fit, with the systematic uncertainties represented by the nuisance parameters in the likelihood with Gaussian priors, are shown in Table~\ref{tab:yields-m} and Fig.~\ref{fig:fit_muons} for the low-\qsq bin and resonant CRs and in Table~\ref{tab:dif_yield} and Figs.~\ref{fig:dif_fit1}--\ref{fig:dif_fit2} for the simultaneous fit. 

\begin{table*}[!htb]
	\begin{center}
	\topcaption{Fit functions used for signal and background sources in each \qsq bin in the muon channel. The \NA{} symbol indicates this background is not included in this region.}
	\begin{tabular}{llll}
	\hline\hline
	Process	 & $\PBp \!\to\! \PKp\MM$ & $\PBp \!\to\! \JPsi(\MM)\PKp$ & $\PBp \!\to\! \PGyP{2S}(\MM)\PKp$ \\
		\hline
		Signal & DCB $+$ Gaussian & Sum of 3 Gaussians & DCB $+$ Gaussian \\
		Comb. \& other \PQb bkg. & Exponential$^\dagger$ & Exponential & Exponential \\
		$\PBp \!\to\! \Kstar\PX$ & DCB ($+$ expon.) & DCB $+$ exponential & DCB $+$ exponential \\
		$\PBp \!\to\! \PGpp\PX$ & DCB & DCB &  DCB \\
		\BKJpmm  & DCB (nearby \qsq)& \NA &  \NA \\
		\BKPsimm & DCB (nearby \qsq)& \NA &  \NA \\
	\hline\hline
	\label{tab:fit_function_muons}	
	\end{tabular}\\
	\end{center}
	\vspace*{-7mm}
	\noindent $^\dagger$ In the last \qsq bin the exponential function is multiplied by $\mKmm - \mmm - m_\PKp$ to account for the phase space suppression.
\end{table*}
\begin{table*}[!htb]
    \centering
    \topcaption{Signal yields in the muon channel in the low-\qsq bin and resonant CRs.}
    \begin{tabular}{lr@{--}lr@{~$\pm$~}l}
    \hline\hline
    \mystrut Channel &  \multicolumn{2}{c}{\qsq range [$\GeVns^2$]}& \multicolumn{2}{c}{Yield} \\
    \hline
    \BKmm & 1.1 & 6.0 & 1267 & 55  \\
    \BKJpmm & 8.41 & 10.24 & 728$\,$000 & 1000 \\
    \BKPsimm & 12.60 &14.44 & 68$\,$300 & 500 \\
     \hline\hline
    \end{tabular}
    \label{tab:yields-m}
\end{table*}

\subsection{Electron channel signal extraction}
\begin{figure*}[htb!]
    \centering
    \includegraphics[width=0.47\textwidth]{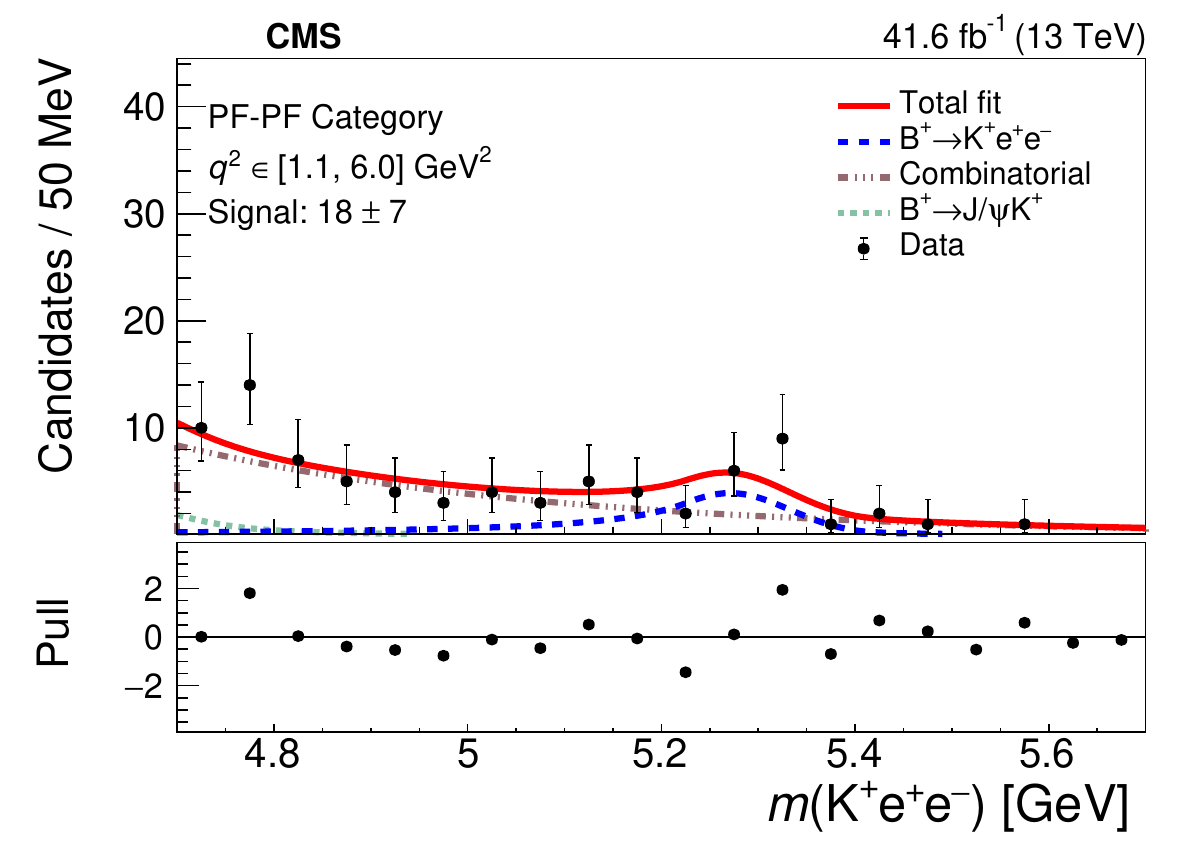} 
    \includegraphics[width=0.47\textwidth]{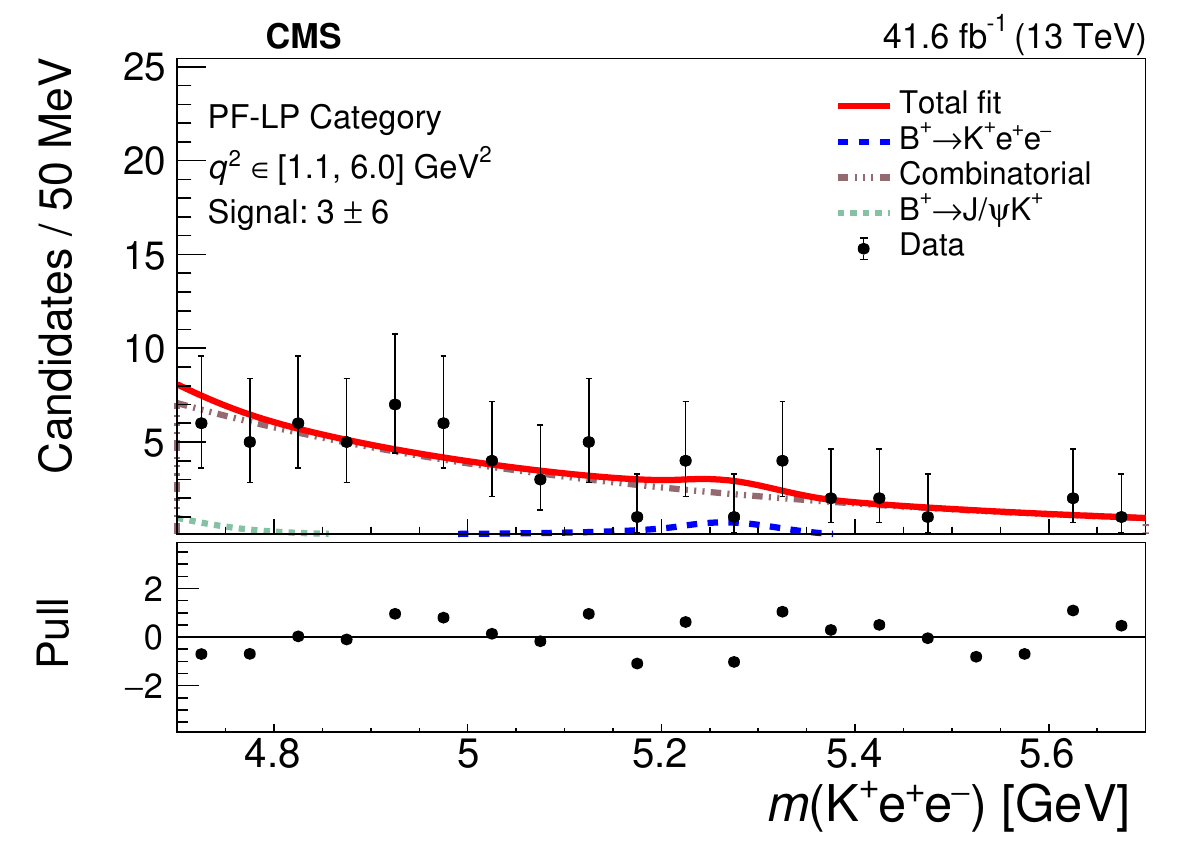}
    \includegraphics[width=0.47\textwidth]{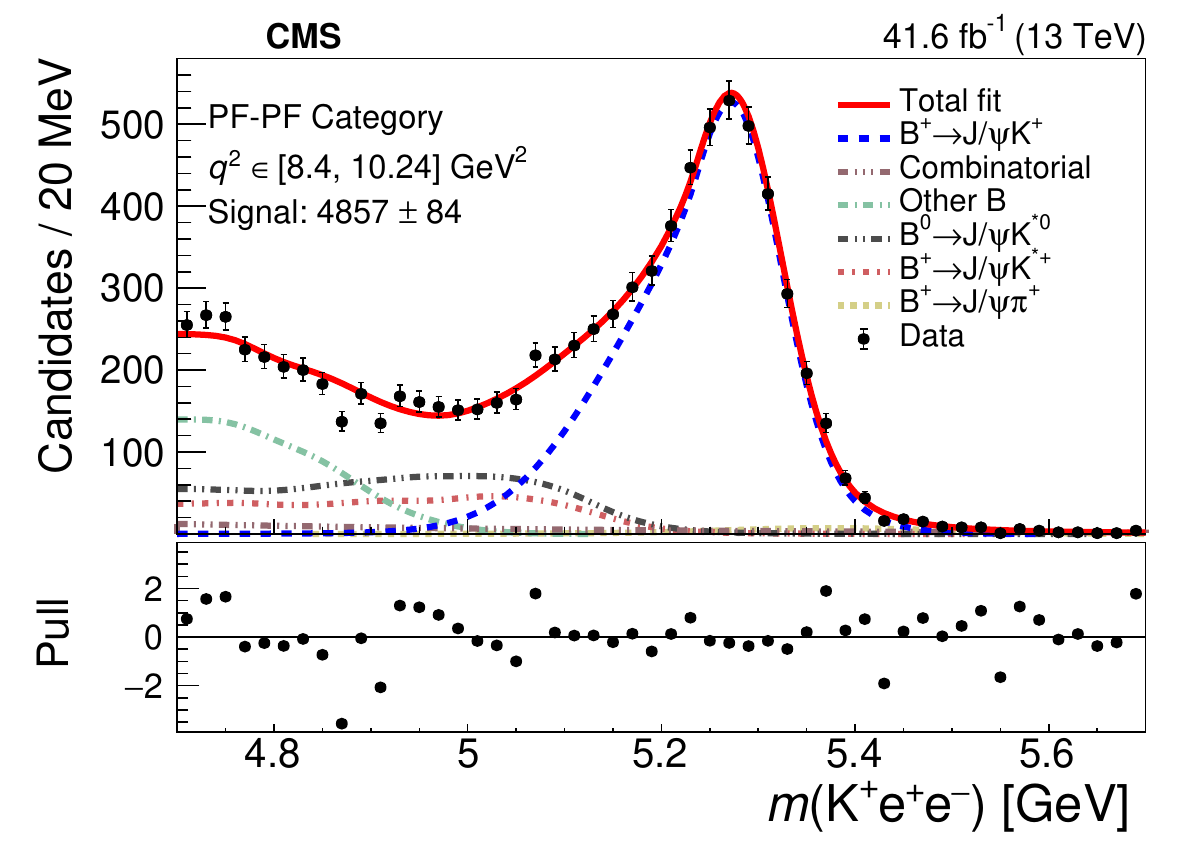} 
    \includegraphics[width=0.47\textwidth]{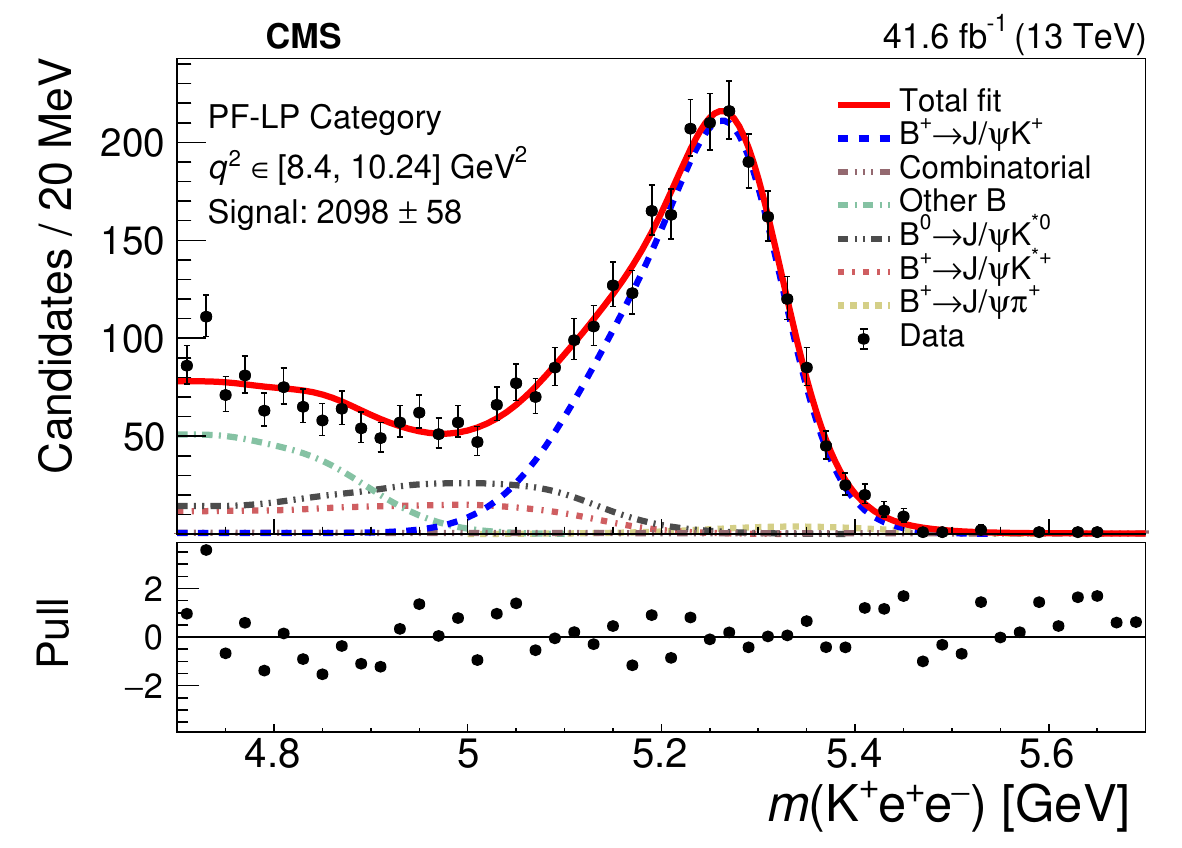}
    \includegraphics[width=0.47\textwidth]{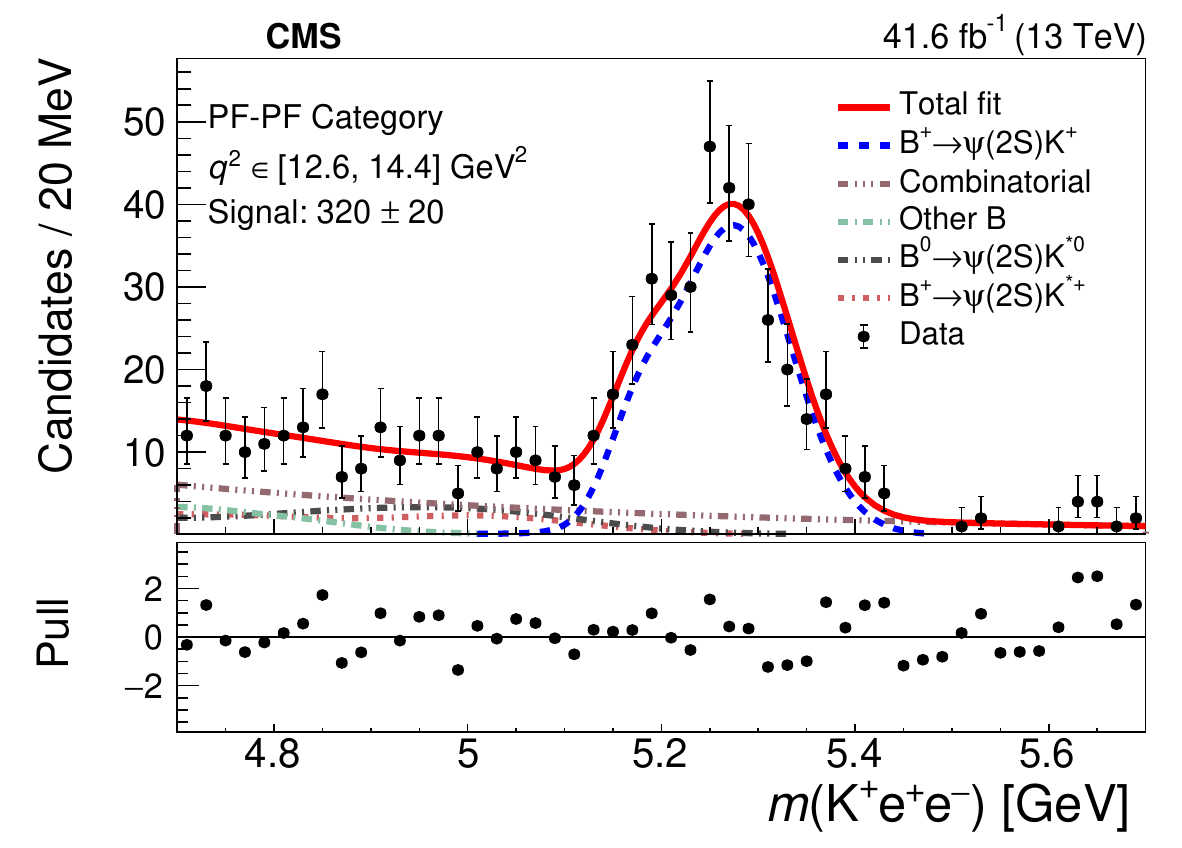}
     \includegraphics[width=0.47\textwidth]{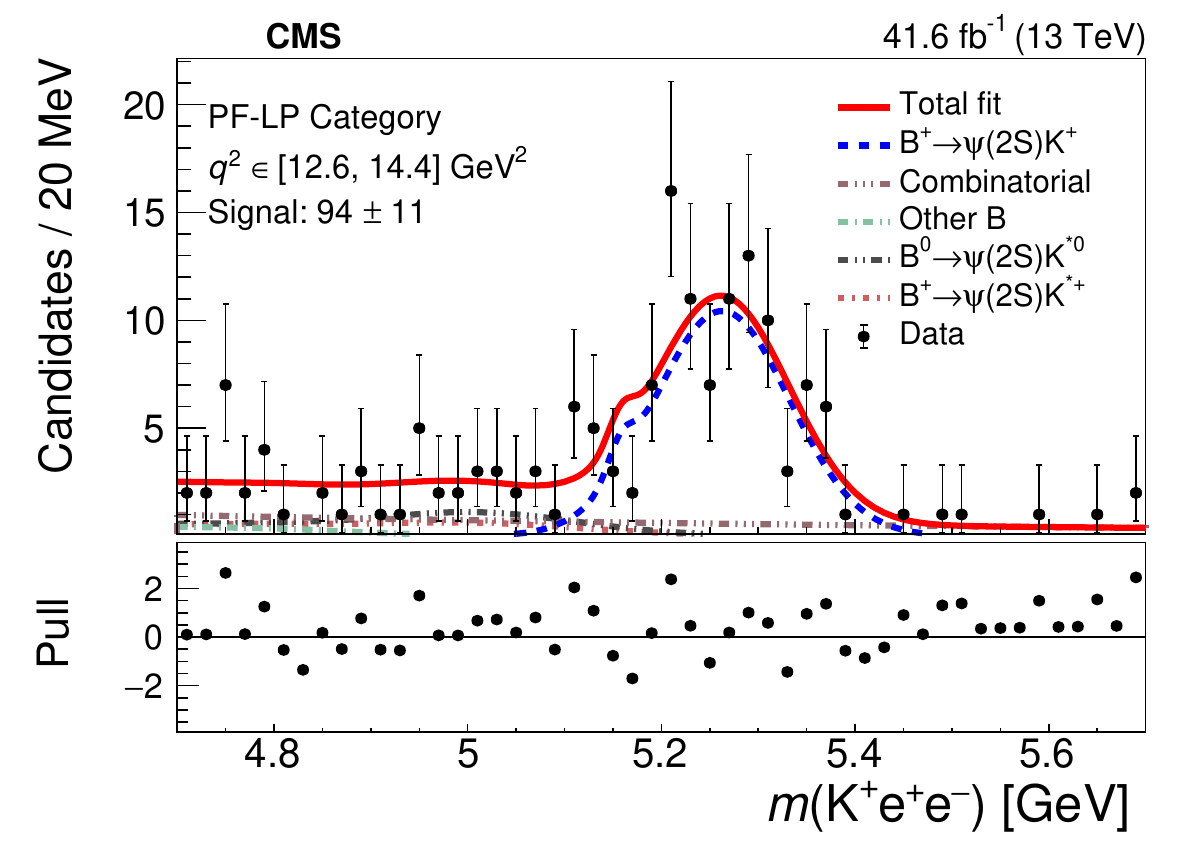} 
    \caption{The $\PKp\EE$ invariant mass spectrum with the results of the fit shown with the red line in the low-\qsq region (upper row), \BKJpee CR (middle row), and \BKPsiee CR (lower row) for the PF-PF (left column) and PF-LP (right column) categories. The shoulder below the nominal \PBp meson mass for the  $\PGyP{2S}$ CR is due to the narrow \qsq range in this bin compared to the size of the radiative tail. Notations are as in Fig.~\protect\ref{fig:fit_muons}.}
    \label{fig:ele_fit}
\end{figure*}

To extract the number of signal events, unbinned maximum likelihood fits to the \PBp candidate invariant mass distribution are performed in three \qsq regions. The \qsq regions considered are: low-\qsq aiming at the signal \BKee decay, and two CRs aiming at \BKJpee and \BKPsiee decays. The high-\qsq bin, above the $\PGyP{2S}$ resonance, has been studied, but the backgrounds in this bin are too large to extract the \BKee signal reliably; consequently, it is not included in the analysis. Fits are done independently in the PF-PF and PF-LP categories. Signal shapes are described by either a DCB function or the sum of a one-sided Crystal Ball (CB) function~\cite{Oreglia} and a Gaussian function. The same background sources as in the muon channel also contribute here, albeit with different relative importance. The partially reconstructed background is similar to that in the muon channel. The combinatorial background is more prominent with respect to the muon channel, because of the higher fraction of hadrons misreconstructed as electrons. Finally, the other \PQb hadron decay background is represented by a separate template. In the electron channel, we use simulated distributions directly to construct various background templates. This is achieved using the kernel density estimator (KDE) method~\cite{kde} applied to simulated events. The only exception is the combinatorial background, for which a simple exponential with the slope and normalization floating in the fit is used. The Cabibbo-suppressed decay background is negligible for the low-\qsq and $\PGyP{2S}$ bins, and therefore is accounted for only in the \BKJpee CR\@. The relative normalization of this background to the signal is fixed from simulation. The shapes of various backgrounds are fixed from simulation and the normalizations are free parameters of the fit, unless specified otherwise. The functions used in the fit are presented in Table~\ref{tab:fit_function_electrons}. Figure~\ref{fig:ele_fit} shows the $\PKp\EE$ invariant mass distribution for the electron channels in the three \qsq regions for the PF-PF and PF-LP categories. The signal yields in these regions are listed in Table~\ref{tab:yields-e}.

\begin{table*}[!htb]
  \centering
  \topcaption{Fit functions used to describe signal and various background components for the electron channel. The \NA{} symbol indicates this background is not included in this region.}
	\begin{tabular}{llll}
\hline\hline
   Process & \BKee & \BKJpee & \BKPsiee \\
    \hline
    Signal & DCB function &  CB $+$ Gaussian & CB + Gaussian \\
    Comb. background & Exponential & Exponential & Exponential \\
    $\PBp \!\to\!\Kstar\PX$ & \NA & KDE template & KDE template \\
    $\PBp \!\to\! \PGpp\PX$ & \NA & CB function &  \NA \\
    \BKJpee  & KDE template &  \NA  & \NA\\
    Other \PQb decays & \NA & KDE template & KDE template \\
    \hline\hline
  \end{tabular}
  \label{tab:fit_function_electrons}  
\end{table*}
\begin{table*}[!htb]
    \centering
    \topcaption{Signal yields in the electron channel in the low-\qsq bin and resonant CRs.}
    \begin{tabular}{lr@{--}lr@{~$\pm$~}lr@{~$\pm$~}l}
    \hline\hline
    \mystrut Channel &  \multicolumn{2}{c}{\qsq range [$\GeVns^2$]}& \multicolumn{2}{c}{PF-PF yield} & \multicolumn{2}{c}{PF-LP yield} \\
    \hline
    \BKee & 1.1 & 6.0 &  17.9 & 7.2 & 3.0 & 5.9 \\
    \BKJpee & 8.41 & 10.24  & 4857 & 84 & 2098 & 58 \\
    \BKPsiee & 12.60 &14.44 & 320 & 20 & 94 & 11 \\
     \hline\hline
    \end{tabular}
    \label{tab:yields-e}
\end{table*}

\section{Systematic uncertainties\label{sec:syst}} 
Statistical uncertainties in the signal yields are propagated into the measurement of \RK. The final statistical uncertainty in \RK is dominated by the yield in the low-\qsq bin in the electron channel, which is of order of 40\% (as can be seen in Fig.~\ref{fig:ele_fit}). 

Systematic uncertainties in this analysis fall into two categories: those due to the finite size of the signal MC samples used for the \Aeps estimates, which are statistical in nature, and those that reflect certain assumptions made in the analysis. Systematic uncertainties are calculated independently for the muon and electron channel parts of \RK, because they largely cancel in the single ratios, $\BKll/\BKJpll$, in each of the two channels. Consequently, only the uncertainties that do not cancel or cancel partially are discussed in this section. The uncertainties in the muon and electron channels are treated as uncorrelated, which is supported by the fact that the two channels have different sources of systematic uncertainties due to differences in the trigger, kinematics (because of the tag-side vs. probe-side selection), and lepton reconstruction. To evaluate the impact of a systematic source, the \RK ratio is remeasured after changing the central value for the source under study by ${\pm}1\sigma$. The difference between the modified and nominal values of the single ratio is used as the systematic uncertainty.

\subsection{Systematic uncertainties in the muon channel single ratio}
The dominant systematic uncertainties in the single ratio $\BKmm/\BKJpmm$, in order of importance, are as follows:
\begin{itemize}
    \item Parameterization of the background function of the \BKmm fit:
     This uncertainty is estimated by using a falling tail of a Gaussian function instead of an exponential function to describe the dominant combinatorial plus other \PB meson decays background. The effect is 1.8\%.
    \item Operating below the trigger plateau: The uncertainty in the trigger turn-on effects was estimated by repeating the measurement after tightening the offline requirements. The corresponding uncertainty is 1.3\%.
	\item Uncertainty in the FONLL scale factors: Simulated events are reweighted in \pt and rapidity of the \PB candidate according to the FONLL predictions. The corresponding uncertainty is 0.9\%.
	\item Parameterization of the background function in the \BKJpmm CR fit: The uncertainty is estimated the same way as in the SR. The effect is 0.6\%.
	\item Description of the \JPsi meson radiative tail: Events in the \BKJpmm channel are selected using a fixed \qsq window. A mismodeling of the tail can lead to an incorrect calculation of efficiency. This effect is estimated by repeating the measurement in a larger \qsq window, by relaxing the lower boundary from 2.9 to 2.8\GeV. The effect is 0.5\%.
	\item Pileup: The pileup profile is reweighted using a ${\pm}4.6\%$ variation in the total inelastic $\Pp\Pp$ cross section~\cite{CMS:2018mlc}. The corresponding uncertainty is 0.4\%.
	\item Parameterization of the signal shape in the low-\qsq \BKmm fit: The uncertainty due to the signal description is estimated by adding an extra Gaussian function to the signal template and repeating the fit. The corresponding systematic uncertainty is 0.3\%.
	\item Uncertainty in the trigger scale factors: The combined statistical and systematic uncertainty from the SF measurement is obtained using the tag-and-probe method. The effect is 0.2\%.
    \item Parameterization of the signal shape in the \BKJpmm fit: This uncertainty is evaluated in the same way as in the low-\qsq bin. The corresponding uncertainty is 0.1\%.
    \item The nonresonant \BKmm contribution in the \BKJpmm CR: The predicted \BKmm contribution in the \JPsi \qsq range using \PYTHIA is subtracted from the signal yield. The effect is 0.1\%. 
 \end{itemize}

Other uncertainty sources were also studied (e.g., several SFs related to the BDT), but found to have a negligible effect ($<$0.01\%) and are therefore omitted. The systematic sources are treated as uncorrelated, and the overall uncertainty in the low-\qsq bin is found to be 2.6\%. The statistical uncertainty in the same bin is 7.5\% and the uncertainty from the limited size of the simulation sample used for the \Aeps calculations is 1.7\%. The uncertainties in the \BKmm/\BKJpmm ratio measurement are summarized in Table~\ref{tab:syst-muon}. For the differential branching fraction measurement the same systematic sources are evaluated in each \qsq bin and the resulting uncertainties are shown in Fig.~\ref{fig:dif_syst}. In most \qsq bins, the data statistical uncertainty is dominant.

\begin{table*}[hbt]
	\centering
        \topcaption{Major sources of uncertainty in the \BKmm/\BKJpmm ratio measurement. \label{tab:syst-muon}}	
	\begin{tabular}{lc}
		\hline\hline
		Source & Impact on the \RK ratio [\%]\\
		\hline
		\mystrut Background description, low-\qsq  bin 	& 1.8\\
		Trigger turn-on 						& 1.3\\
		Reweighting in \pt and rapidity			& 0.9\\
		Background description, \JPsi CR		& 0.6\\
		\JPsi meson radiative tail description		& 0.5\\
		Pileup							& 0.4\\
		Signal shape description				& 0.3\\
		Trigger efficiency					& 0.2\\
		\JPsi resonance shape  description		& 0.1\\
		Nonresonant contribution to the \JPsi CR	& 0.1\\
		\hline
		Total	systematic uncertainty			& 2.6\\
		\hline
		Statistical uncertainty in MC samples		& 1.7\\
		Statistical uncertainty in data			& 7.5\\
		\hline
		Total uncertainty					& 8.1\\
		\hline\hline
	\end{tabular}
\end{table*}

\subsection{Systematic uncertainties in the electron channel single ratio}
Since the yield of \BKee events in the low-\qsq bin is small, the overall uncertainty is dominated by the statistical uncertainty. The sources of systematic uncertainties in the electron channel are generally different from those in the muon channel, due to different kinematics and reconstruction performance. 

The dominant uncertainties in the single ratio \BKee/\BKJpee, in order of importance, are as follows:
\begin{itemize} 
	\item Parameterization of signal and background shapes: Both the signal and background shapes in the fits of the low-\qsq SR and \JPsi CR are modified and the difference is taken as an uncertainty. The modification is done by replacing the KDE-based templates with analytical functions similar to the ones used in the muon channel. The signal description in the low-\qsq bin was modified by adding an extra Gaussian function, and the exponential describing the combinatorial background was replaced by a second-order Chebyshev polynomial. The combined uncertainty in the fit parameterization because of the signal and background variations is about 5\% each in the PF-PF and PF-LP categories.
	\item Constraint on the \BKJpee contribution: The normalization of the \JPsi meson leakage background in the \BKee channel in the low-\qsq region is constrained to the expected yield from the fit in the \BKJpee CR\@. The uncertainty is dominated by the statistical uncertainty in the yield of this background and is estimated using pseudo-experiments, while the contribution from the radiative tail mismodeling is negligible. The uncertainty amounts to 4 (9)\% in the PF-PF (PF-LP) category.
	\item The BDT efficiency stability: The BDT WP is chosen to maximize the expected significance of the \BKee signal in the low-\qsq region. A variation of the WP corresponding to a variation of $\pm 10\%$ in the expected significance is chosen to evaluate the stability. The corresponding systematic uncertainty is 2 (5)\% for the PF-PF (PF-LP) category.
	\item The BDT cross-validation: To allow for the entire data set to be used both in the BDT training and testing, an eight-fold cross validation is used, resulting in eight different BDTs. The spread in the efficiency of these BDTs is used as the uncertainty, amounting to 2 (3)\% in the PF-PF (PF-LP) category.
	 \item Triggers in the \PB parking data set: In the electron channel all the \PB parking triggers are used for the measurement. The small kinematical correlation between the tag and probe sides could have an impact on the \RK value. Due to a complicated mixture of L1 and HLT triggers, some of which are prescaled as a function of the instantaneous luminosity, the composition in terms of trigger paths is different in simulated samples than in data. To account for this, the trigger efficiency ratio between the low-\qsq SR and \JPsi CR is estimated for several individual trigger paths and the variation of this ratio is taken as a systematic uncertainty amounting to 1 (4)\% for the PF-PF (PF-LP) category.
    \item The BDT scale factor: This SF accounts for the differences in the BDT WP efficiency between data and simulation. The efficiency of the nominal WP with respect to a loose selection, which is nearly fully efficient for signal, is computed in the \JPsi CR, and transferred to the low-\qsq bin using simulation. The uncertainty from this source is 1 (2)\%  in the PF-PF (PF-LP) category.  
	\item Description of the \JPsi meson radiative tail: This uncertainty is estimated in the same way as in the muon channel, except that the enlargement of the \qsq window is bigger in the case of electrons to account for the larger radiative tail. The effect is 1\% for both PF-PF and PF-LP categories.
 \end{itemize} 
Other sources of uncertainty have an impact of less than 1\% on the \BKee/\BKJpee ratio. The reason the uncertainties in the BDT performance are much larger in the electron channel than in the muon one is the much tighter WP used, which results in effectively smaller training samples and higher sensitivity to data-to-simulation differences in the tails of the distributions of the various input variables. All sources of systematic uncertainties are treated as uncorrelated, which results in an overall systematic uncertainty in the \BKee/\BKJpee single ratio of 7 and 13\% for the PF-PF and PF-LP categories, respectively. The systematic uncertainties are small compared to the statistical ones which are 40 (200)\% for the PF-PF (PF-LP) channel. Table~\ref{tab:syst-electron} summarizes the uncertainties in the electron channel.

\begin{table*}[hbt]
	\centering
        \topcaption{Major sources of uncertainty in the \BKee/\BKJpee ratio measurement in the PF-PF and PF-LP categories. The last row shows the statistical uncertainty, which is the same as the total uncertainty within the quoted precision.
         \label{tab:syst-electron}}	
	\begin{tabular}{lcc}
		\hline\hline
		Source & \multicolumn{2}{c}{Impact on the \RK ratio [\%]}\\
			    & PF-PF & PF-LP \\
		\hline
		Signal and background description	 	& 5 & 5\\
		\JPsi event leakage to the low-\qsq bin	& 4 & 9\\
		BDT efficiency stability				& 2 & 5\\
		BDT cross validation					& 2 & 3\\
		Trigger efficiency					& 1 & 4\\
		BDT data/simulation difference			& 1 & 2\\
		\JPsi meson radiative tail description		& 1 & 1\\
		\hline
		Total systematic uncertainty			& 7 & 13 \\
		\hline
		Statistical and total uncertainty			& 40 & 200\\
		\hline\hline
	\end{tabular}
\end{table*}

\section{Results\label{sec:results}}
In this section, we report the following results:
\begin{itemize}
	\item measurement of the differential branching fraction of the \BKmm decay in the full \qsq range, excluding the \JPsi and $\PGyP{2S}$ resonances;
	\item measurement of the integrated branching fractions of the \BKmm decay in the low-\qsq region and in the full \qsq range; and
	\item measurement of \RK in the low-\qsq region.
\end{itemize} 

To validate the analysis procedure we performed several cross-checks, the two most important ones being the measurement of the \Rpsi and \RJPsi ratios. The former is defined by Eq.~(\ref{eq:Rpsi2S}), \ie, exchanging the \BKll decay by the \BKPsill one. The latter is the ratio of the \BKJpmm and \BKJpee branching fractions. Both these ratios are expected and measured~\cite{Workman:2022ynf} to be flavor-universal (\ie, equal to unity), with high precision. An important added value of the \RJPsi cross-check is that various efficiencies and corrections that cancel out in the \RK double ratio, only partially cancel or do not cancel in the \RJPsi single ratio. Therefore, this cross check also validates the systematic uncertainties that cancel in the \RK ratio.

The results of our measurement in the combination of the PF-PF and PF-LP categories are: $\Rpsi = 0.966^{+0.071}_{-0.066}$ and $\RJPsi = 1.006^{+0.020}_{-0.019}$, where the uncertainties shown are statistical only. Both the \Rpsi and \RJPsi measurements are consistent with unity within one standard deviation, with a precision of a few percent. The systematic uncertainty due to the lack of cancellation of various efficiencies and from sources discussed in Section~\ref{sec:syst} for both ratios is estimated to be around 7\%. Consistent results are obtained in the PF-PF and PF-LP categories separately, albeit with larger statistical uncertainties.

\subsection{Measurement of the differential \texorpdfstring{\BR(\PBp $\to\PKp$ {\boldmath $\mu^+\mu^-$})}{Br(B+ to K+ mu+ mu-)}}
To reduce the systematic uncertainty, the differential branching fraction in each \qsq bin is normalized to $\BR(\BKJpmm)$,
\begin{linenomath}
\ifthenelse{\boolean{cms@external}}
{ 
\begin{multline}
  \BR(\BKmm)[\qsqmin,\qsqmax]  = \\ \frac{N_{\BKmm}[\qsqmin,\qsqmax]}{N_{\BKJpmm}[8.41,10.24]\GeV^2} \\
  \times  \frac{(\Aepst)_{\BKJpmm}[8.41,10.24]\GeV^2}{(\Aepst)_{\BKmm}[\qsqmin,\qsqmax]} \\
  \times \BR(\BKJp)\BR(\JPsi \!\to\! \MM),
\label{eq:brlow_form}
\end{multline}
} 
{ 
\begin{multline}
  \BR(\BKmm)[\qsqmin,\qsqmax]  =  \frac{N_{\BKmm}[\qsqmin,\qsqmax]}{N_{\BKJpmm}[8.41,10.24]\GeV^2} \\
     \times  \frac{(\Aepst)_{\BKJpmm}[8.41,10.24]\GeV^2} {(\Aepst)_{\BKmm}[\qsqmin,\qsqmax]} \BR(\BKJp)\BR(\JPsi \!\to\! \MM),
\label{eq:brlow_form}
\end{multline}
}
\end{linenomath}
where $N_{\BKmm}$ and $N_{\BKJpmm}$ are the measured yields from the fit in the \qsq region indicated in the brackets and $\BR(\BKJpmm)$ is the world-average value of the \BKJpmm branching fraction~\cite{Workman:2022ynf}. The measured differential branching fraction of the \BKmm decay is summarized in Table~\ref{tab:diff_BF}. The correlation matrix between the extracted values of the differential branching fraction in different \qsq bins is shown in Fig.~\ref{fig:dif_cor}.

\begin{figure}[!htb]
	\centering
		\includegraphics[width=0.5\textwidth]{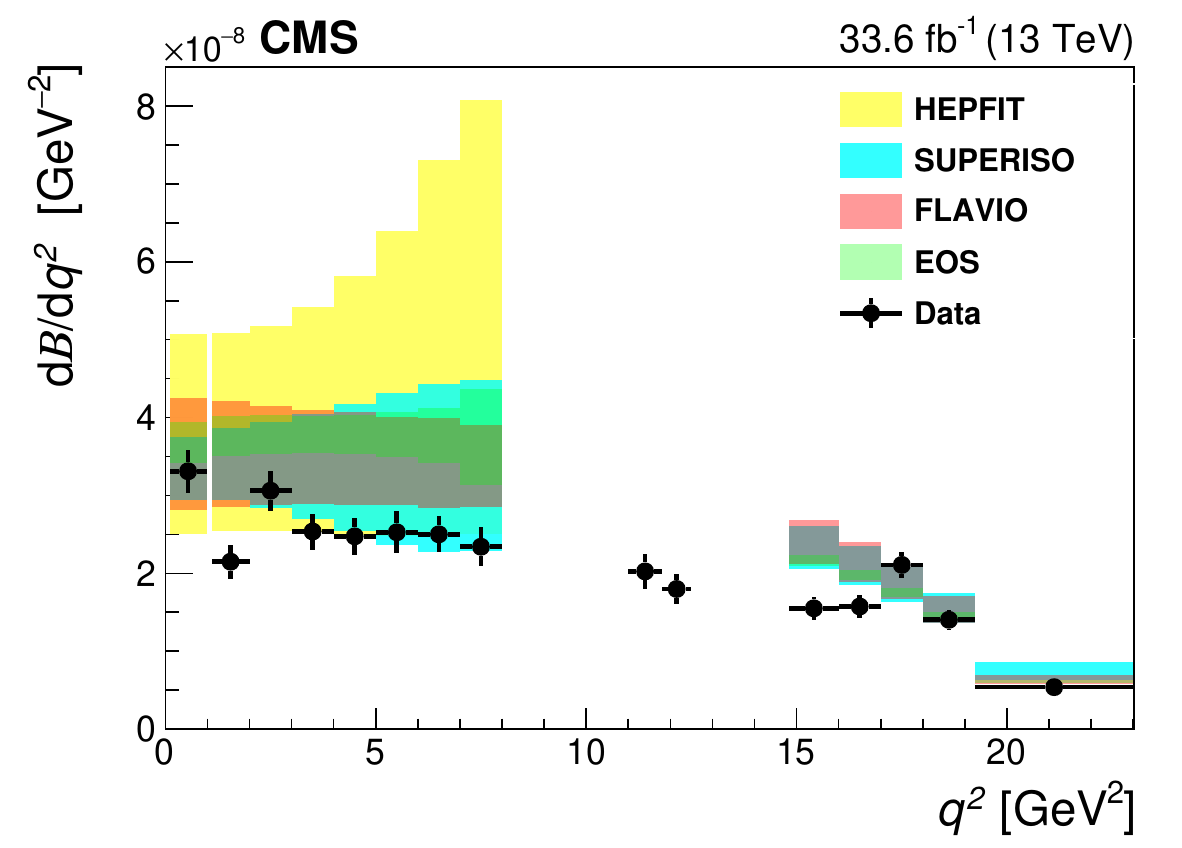} 		
	\caption{Comparison of the measured differential \BKmm branching fraction with the theoretical predictions obtained using \textsc{hepfit}, \textsc{superIso},  \textsc{flavio}, and \textsc{eos} packages. The \textsc{hepfit} predictions are available only for $\qsq < 8\GeV^2$.
		\label{fig:diff_theory}}
\end{figure}

Since there are various and somewhat different theoretical predictions for the SM value of $\ddinline{\BR(\BKmm)}{\qsq}$, we compare our measurement with the predictions from the \textsc{hepfit}~\cite{DeBlas:2019ehy,Alasfar:2020mne,Ciuchini:2021smi,Ciuchini:2022wbq} v1.0, \textsc{superiso}~\cite{Mahmoudi:2007vz,Mahmoudi:2008tp} v4.1, \textsc{flavio}~\cite{Straub:2018kue} v2.5.5, and \textsc{eos}~\cite{EOSAuthors:2021xpv,Gubernari:2022hxn} v1.0.8 packages. These models rely on different approaches in evaluating the effects of nonlocal form-factor contributions, which is reflected in a sizable difference between the uncertainties in their predictions. None of the calculations can reliably describe the regions between the \JPsi and $\PGyP{2S}$ resonances; hence the predictions are not shown in this \qsq range. In addition, the \textsc{hepfit} package only gives predictions for $\qsq < 8\GeV^2$. The comparison of our measurement with the prediction of these  models is shown in Fig.~\ref{fig:diff_theory}. The measured differential branching fraction of the \BKmm decay is generally lower than the theoretical predictions for $\qsq < 17\GeV^2$, which is consistent with the results reported by LHCb~\cite{LHCb:2012bin,LHCb:2014cxe}.

\begin{table}[!htb]
	\centering
	\topcaption{The  \BKmm  branching fraction, $\ddinline{\BR(\BKmm)}{\qsq}$ integrated over the specified \qsq
range, for the individual \qsq bins.. The uncertainties in the yields are statistical uncertainties from the fit, while the branching fraction uncertainties include both the statistical and systematic components.\label{tab:diff_BF}}
	\label{tab:dif_yield}
	\begin{tabular}{r@{--}lcc}
		\hline\hline
		\multicolumn{2}{c}{\mystrut\qsq range} & Signal yield & Branching fraction\\
		\multicolumn{2}{c}{[$\GeVns^2$]} &   & [$10^{-8}$]\\
		\hline
		 0.1	& 	0.98 		&	$260 \pm 20$	& $2.91 \pm 0.24$\\
		 1.1	&	2.0 		&	$197 \pm 19$	& $1.93 \pm 0.20$\\
		 2.0	&	3.0 		&	$306 \pm 23$	& $3.06 \pm 0.25$\\
		 3.0	&	4.0 		&	$260 \pm 21$	& $2.54 \pm 0.23$\\
		 4.0	&	5.0 		&	$251 \pm 23$	& $2.47 \pm 0.24$\\
		 5.0	&	6.0 		&	$264 \pm 27$	& $2.53 \pm 0.27$\\
		 6.0	&	7.0 		& 	$267 \pm 21$	& $2.50 \pm 0.23$\\
		 7.0	&	8.0 		& 	$256 \pm 23$	& $2.34 \pm 0.25$\\
		 11.0	&	11.8  	        &       $207 \pm 19$	& $1.62 \pm 0.18$\\
		 11.8	&	12.5 		& 	$172 \pm 16$	& $1.26 \pm 0.14$\\
		 14.82  &	16.0 		& 	$272 \pm 20$	& $1.83 \pm 0.17$\\
		 16.0	&	17.0 		& 	$246 \pm 17$	& $1.57 \pm 0.15$\\
		 17.0	&	18.0 		& 	$317 \pm 19$	& $2.11 \pm 0.16$\\
		 18.0	&	19.24 	        & 	$242 \pm 19$	& $1.74 \pm 0.15$\\
		 19.24  &	22.9 		& 	$158 \pm 19$	& $2.02 \pm 0.30$\\
		  \hline\hline
	\end{tabular}
\end{table}

\subsection{Measurement of the integrated \texorpdfstring{\BR(\PBp $\to\PKp${\boldmath $\mu^+\mu^-$})}{Br(B+ to K+ mu+ mu-)}}
The low-\qsq region is especially interesting because it is not affected by higher-mass resonances and their interference effects. Consequently, the branching fraction of the \BKmm decay in this \qsq region can be measured with relatively little theoretical dependence, using Eq.~(\ref{eq:brlow_form}) over the $[1.1,6.0]\GeV^2$ \qsq range. In this range, the acceptance times efficiency is essentially independent of the \qsq value, as shown in Fig.~\ref{fig:acceff_muon}. Therefore, any dependence on theory can only arise from residual differences in the kinematic distributions of the final-state particles between theory and data. Since the acceptance times efficiency is evaluated using events generated with the \EVTGEN \BKmm model (discussed in Section~\ref{sec:MC}), which has been tuned using $\Pep\Pem$ experimental data, as well as lattice calculations, any such residual differences are expected to be minimal. The resulting measurement is
\begin{linenomath}
\ifthenelse{\boolean{cms@external}}
{ 
\begin{multline}
  \BR(\BKmm)[1.1,6.0]\GeV^2 = \\
  (12.42 \pm 0.54\stat \pm 0.11\,\text{(MC~stat)} \pm 0.40\syst)  \times  10^{-8} \\
  = (12.42 \pm 0.68) \times 10^{-8}.
\label{eq:brlow}
\end{multline}
} 
{ 
\begin{multline}
  \BR(\BKmm)[1.1,6.0]\GeV^2  \\
  =  (12.42 \pm 0.54\stat \pm 0.11\,\text{(MC~stat)} \pm 0.40\syst)  \times  10^{-8}  = (12.42 \pm 0.68) \times 10^{-8}.
\label{eq:brlow}
\end{multline}
}
\end{linenomath}
This result is consistent with the present world-average value of $(12.6 \pm 1.2) \times 10^{-8}$~\cite{Workman:2022ynf} in a very similar range, $1.0 < \qsq < 6.0\GeV^2$, and has a 40\% smaller uncertainty. It is also consistent with and has a similar uncertainty as the LHCb measurement~\cite{LHCb:2014cxe} in the $1.1 < \qsq < 6.0\GeV^2$ range, $(11.86 \pm 0.68) \times 10^{-8}$, which presently dominates the world-average value. (The larger uncertainty in the world-average value is due to a scale factor of 1.9 introduced to address the tension between the individual results.) The comparison between our measurement and the theoretical predictions described above is shown in Table~\ref{tab:total_lowq2_bf}. All of the theoretical estimates are higher than our measurement in the low-\qsq region.

In order to determine the integrated $\BR(\BKmm)$, the result of Eq.~(\ref{eq:brlow}) must be divided by the fraction of events in the low-\qsq bin. This fraction cannot be taken directly from data because of the interference effects and resonant contributions. Two theoretical models are used to obtain the differential branching fraction distribution in the full \qsq range, based on the \textsc{flavio} and \textsc{superiso} packages. The resulting $\BR(\BKmm)$ integrated branching fractions are $43.5 \pm 1.9\stat \pm 0.4\,\text{(MC~stat)} \pm 1.4\syst = 43.5 \pm 2.4$  (\textsc{flavio}) and $43.9 \pm 1.9\stat \pm 0.4\,\text{(MC~stat)} \pm 1.4\syst = 43.9 \pm 2.4$ (\textsc{superiso}), where the theoretical model uncertainty is not included. Both results are in good agreement with the world-average value of $(45.3 \pm 3.5)\times 10^{-8}$~\cite{Workman:2022ynf}, as well as with the LHCb measurement of $(43.7 \pm 2.7)\times 10^{-8}$ that explicitly subtracts various resonant contributions~\cite{LHCb:2016due}. For the calculation of the integrated branching fraction $\BR(\BKmm)$, only \textsc{flavio} and \textsc{superiso} are used because \textsc{eos} and \textsc{hepfit} do not provide predictions for the entire \qsq range.

\begin{table}[!htb]
        \centering
        \topcaption{Comparison of the $\BR(\BKmm)$ branching fraction measurement in the low-\qsq range and the theoretical predictions based on the \textsc{eos}, \textsc{flavio}, \textsc{superiso}, and \textsc{hepfit} packages.\label{tab:total_lowq2_bf}}
        \begin{tabular}{lc}
                \hline\hline
                \mystrut Source & $\BR(\BKmm)[1.1,6.0]\GeV^2$ \\
                & [$10^{-8}$]  \\
                \hline
                Measurement    & $ 12.42 \pm 0.68$ \\
                \textsc{eos}    & $ 18.9 \pm 1.3$ \\
                \textsc{flavio} & $ 17.1 \pm 2.7$    \\
                \textsc{superiso}    & $ 16.5 \pm 3.4$ \\
                \textsc{hepfit}    & $ 19.8 \pm 7.3$ \\
                \hline\hline
        \end{tabular}
\end{table}

\subsection{Measurement of \texorpdfstring{\RK}{R(K)}}
For the \RK measurement, both the \BKmm and \BKee channels are used in the low-\qsq region, along with the \BKJpmm and \BKJpee CRs, separately in the PF-PF and PF-LP electron channel categories and in their combination,. The details are given in Appendix~\ref{sec:formalism}. The systematic uncertainties are incorporated into the likelihood function assuming Gaussian distributions. The correlation of a few minor uncertainties between the two channels were found to have a negligible effect on the results. A profile likelihood is used to obtain the confidence interval of the parameter of interest, \iRK.

\begin{figure}[!htb]
 \centering
    \includegraphics[width=0.5\textwidth]{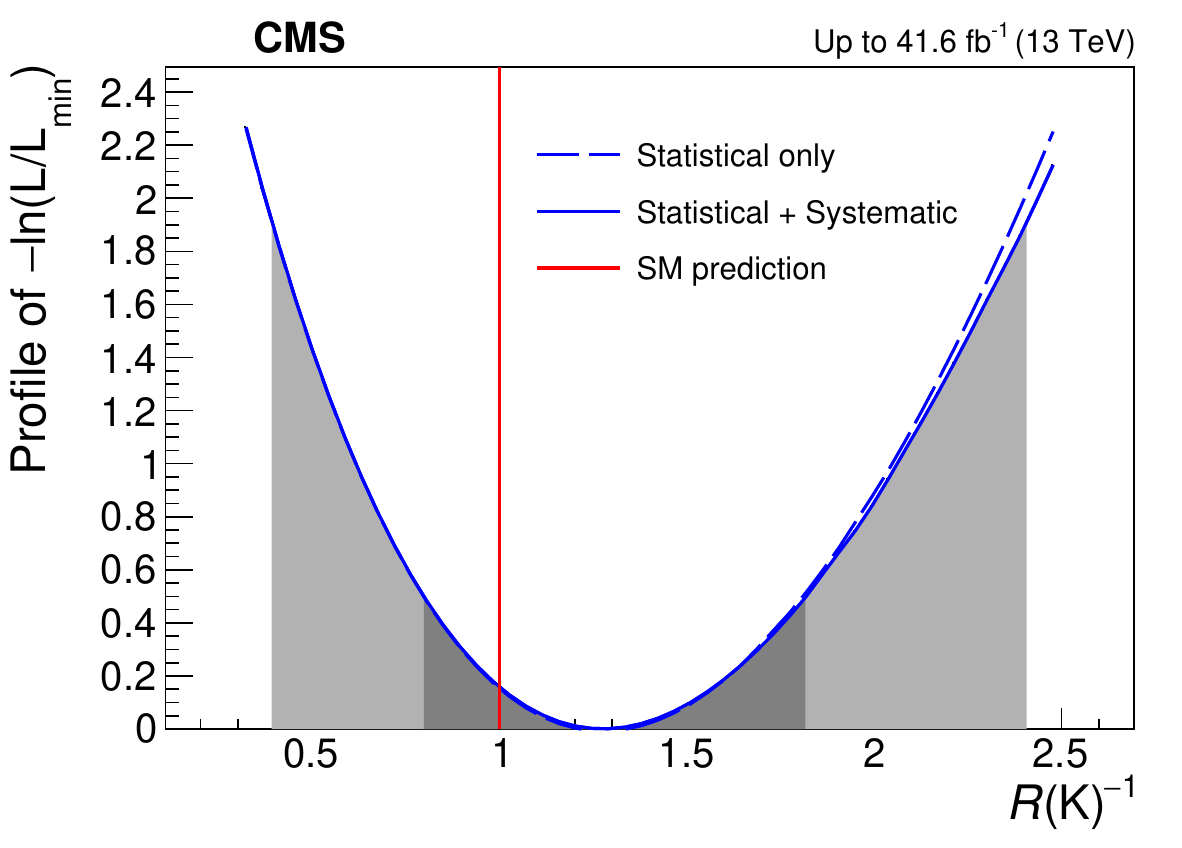}
    \caption{Log likelihood function from the fit profiled as a function of \iRK. The dark and light grey area indicates the $\pm$ 1 and $\pm$ 2 $\sigma$ bands respectively.}
    \label{fig:RK_pl}
\end{figure}

The values of \iRK extracted in the PF-PF and PF-LP categories are $1.40^{+0.63}_{-0.56}$ and $0.50^{+1.06}_{-0.88}$, respectively, where the uncertainties are statistical only. These values are consistent with each other and with unity. As expected from the corresponding signal yields, the PF-PF result is much more precise. The maximum likelihood fit for the combination of the PF-PF and PF-LP categories gives $\iRK = 1.28^{+0.53}_{-0.47}$, where the uncertainties are statistical only. The profile log likelihood of the combined fit, $\ln(L/L_{\text{min}})$, as a function of \iRK is shown in Fig.~\ref{fig:RK_pl}. The measurement of \RK and its 68\% confidence interval can be obtained from those for \iRK,
\begin{linenomath}
\begin{equation}
\RK= 0.78^{+0.46}_{-0.23}\stat^{+0.09}_{-0.05}\syst = 0.78^{+0.47}_{-0.23}, 
\end{equation}
\end{linenomath}
which is within one standard deviation from the SM prediction of approximately unity. A summary of the available \RK measurements is shown in Fig.~\ref{fig:summary}. 

\section{Summary\label{sec:summary}}
We have reported the first test of lepton flavor universality with the CMS experiment at the LHC in \BpmKmm and \BpmKee decays, as well as a measurement of differential and integrated branching fractions of the nonresonant \BpmKmm decay. The analysis has been made possible by a dedicated data set of proton-proton collisions at $\sqrt{s} = 13\TeV$ recorded in 2018, using a special high-rate data stream designed for collecting about 10 billion unbiased \PQb hadron decays. The ratio of the branching fractions $\BR(\BpmKmm)$ to $\BR(\BpmKee)$ has been is determined from the measured double ratio \RK of these decays to the respective branching fractions of the \BpmKJp $(\JPsi\!\to\!\MM)$ and $(\JPsi\!\to\!\EE)$ decays, which allow for significant cancellation of systematic uncertainties. The ratio \RK has been measured in the range $1.1 <\qsq< 6.0\GeV^2 $, where $q$ is the invariant mass of the lepton pair, and was found to be $\RK=0.78^{+0.47}_{-0.23}$, in agreement with the standard model expectation of $\approx$1. This measurement is limited by the statistical precision of the electron channel. The integrated branching fraction in the same \qsq range, $\BR(\BpmKmm) =  (12.42 \pm 0.68)\times 10^{-8}$, is consistent with and has a comparable precision to the present world average. This work has demonstrated the flexibility of the CMS trigger and data acquisition system and has paved the way to many other studies of a large unbiased sample of \PQb hadron decays collected by CMS in 2018.

\begin{acknowledgments}
We congratulate our colleagues in the CERN accelerator departments for the excellent performance of the LHC and thank the technical and administrative staffs at CERN and at other CMS institutes for their contributions to the success of the CMS effort. In addition, we gratefully acknowledge the computing centers and personnel of the Worldwide LHC Computing Grid and other centers for delivering so effectively the computing infrastructure essential to our analyses. Finally, we acknowledge the enduring support for the construction and operation of the LHC, the CMS detector, and the supporting computing infrastructure provided by the following funding agencies: SC (Armenia), BMBWF and FWF (Austria); FNRS and FWO (Belgium); CNPq, CAPES, FAPERJ, FAPERGS, and FAPESP (Brazil); MES and BNSF (Bulgaria); CERN; CAS, MoST, and NSFC (China); MINCIENCIAS (Colombia); MSES and CSF (Croatia); RIF (Cyprus); SENESCYT (Ecuador); MoER, ERC PUT and ERDF (Estonia); Academy of Finland, MEC, and HIP (Finland); CEA and CNRS/IN2P3 (France); SRNSF (Georgia); BMBF, DFG, and HGF (Germany); GSRI (Greece); NKFIH (Hungary); DAE and DST (India); IPM (Iran); SFI (Ireland); INFN (Italy); MSIP and NRF (Republic of Korea); MES (Latvia); LAS (Lithuania); MOE and UM (Malaysia); BUAP, CINVESTAV, CONACYT, LNS, SEP, and UASLP-FAI (Mexico); MOS (Montenegro); MBIE (New Zealand); PAEC (Pakistan); MES and NSC (Poland); FCT (Portugal); MESTD (Serbia); MCIN/AEI and PCTI (Spain); MOSTR (Sri Lanka); Swiss Funding Agencies (Switzerland); MST (Taipei); MHESI and NSTDA (Thailand); TUBITAK and TENMAK (Turkey); NASU (Ukraine); STFC (United Kingdom); DOE and NSF (USA).
	
\hyphenation{Rachada-pisek} Individuals have received support from the Marie-Curie program and the European Research Council and Horizon 2020 Grant, contract Nos.\ 675440, 724704, 752730, 758316, 765710, 824093, and COST Action CA16108 (European Union); the Leventis Foundation; the Alfred P.\ Sloan Foundation; the Alexander von Humboldt Foundation; the Science Committee, project no. 22rl-037 (Armenia); the Belgian Federal Science Policy Office; the Fonds pour la Formation \`a la Recherche dans l'Industrie et dans l'Agriculture (FRIA-Belgium); the Agentschap voor Innovatie door Wetenschap en Technologie (IWT-Belgium); the F.R.S.-FNRS and FWO (Belgium) under the ``Excellence of Science -- EOS" -- be.h project n.\ 30820817; the Beijing Municipal Science \& Technology Commission, No. Z191100007219010 and Fundamental Research Funds for the Central Universities (China); the Ministry of Education, Youth and Sports (MEYS) of the Czech Republic; the Shota Rustaveli National Science Foundation, grant FR-22-985 (Georgia); the Deutsche Forschungsgemeinschaft (DFG), under Germany's Excellence Strategy -- EXC 2121 ``Quantum Universe" -- 390833306, and under project number 400140256 - GRK2497; the Hellenic Foundation for Research and Innovation (HFRI), Project Number 2288 (Greece); the Hungarian Academy of Sciences, the New National Excellence Program - \'UNKP, the NKFIH research grants K 124845, K 124850, K 128713, K 128786, K 129058, K 131991, K 133046, K 138136, K 143460, K 143477, 2020-2.2.1-ED-2021-00181, and TKP2021-NKTA-64 (Hungary); the Council of Science and Industrial Research, India; ICSC -- National Research Center for High Performance Computing, Big Data and Quantum Computing, funded by the NextGenerationEU program (Italy); the Latvian Council of Science; the Ministry of Education and Science, project no. 2022/WK/14, and the National Science Center, contracts Opus 2021/41/B/ST2/01369 and 2021/43/B/ST2/01552 (Poland); the Funda\c{c}\~ao para a Ci\^encia e a Tecnologia, grant CEECIND/01334/2018 (Portugal); the National Priorities Research Program by Qatar National Research Fund; MCIN/AEI/10.13039/501100011033, ERDF ``a way of making Europe", and the Programa Estatal de Fomento de la Investigaci{\'o}n Cient{\'i}fica y T{\'e}cnica de Excelencia Mar\'{\i}a de Maeztu, grant MDM-2017-0765 and Programa Severo Ochoa del Principado de Asturias (Spain); the Chulalongkorn Academic into Its 2nd Century Project Advancement Project, and the National Science, Research and Innovation Fund via the Program Management Unit for Human Resources \& Institutional Development, Research and Innovation, grant B37G660013 (Thailand); the Kavli Foundation; the Nvidia Corporation; the SuperMicro Corporation; the Welch Foundation, contract C-1845; and the Weston Havens Foundation (USA)	
\end{acknowledgments}

\bibliography{auto_generated} 
\clearpage
\appendix
\numberwithin{figure}{section}
\numberwithin{table}{section}
\section{Appendix}
\label{sec:appendix}
\subsection{Additional figures}
In this section, we include auxiliary figures that are omitted from the main body of the paper.
\begin{figure}[htb]
	\centering
    \includegraphics[width=0.45\textwidth]{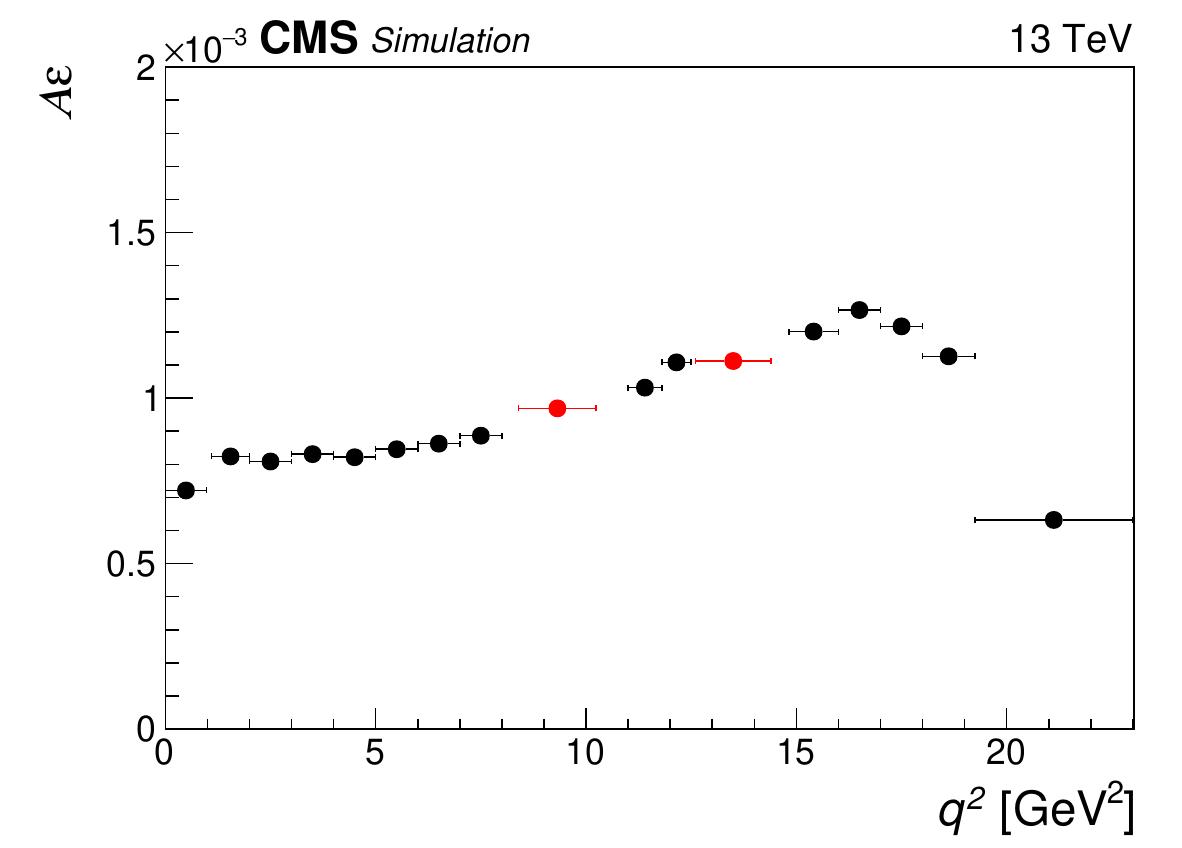}
	\caption{The product of acceptance and efficiency ($\Aeps$) of the \BKmm channel, as a function of the muon pair \qsq, as measured in simulated signal events, after all the corrections applied. Regions corresponding to resonances are displayed with red markers.
	\label{fig:acceff_muon}}
\end{figure}
\begin{figure}[htb]
	\centering
	\includegraphics[width=0.45\textwidth]{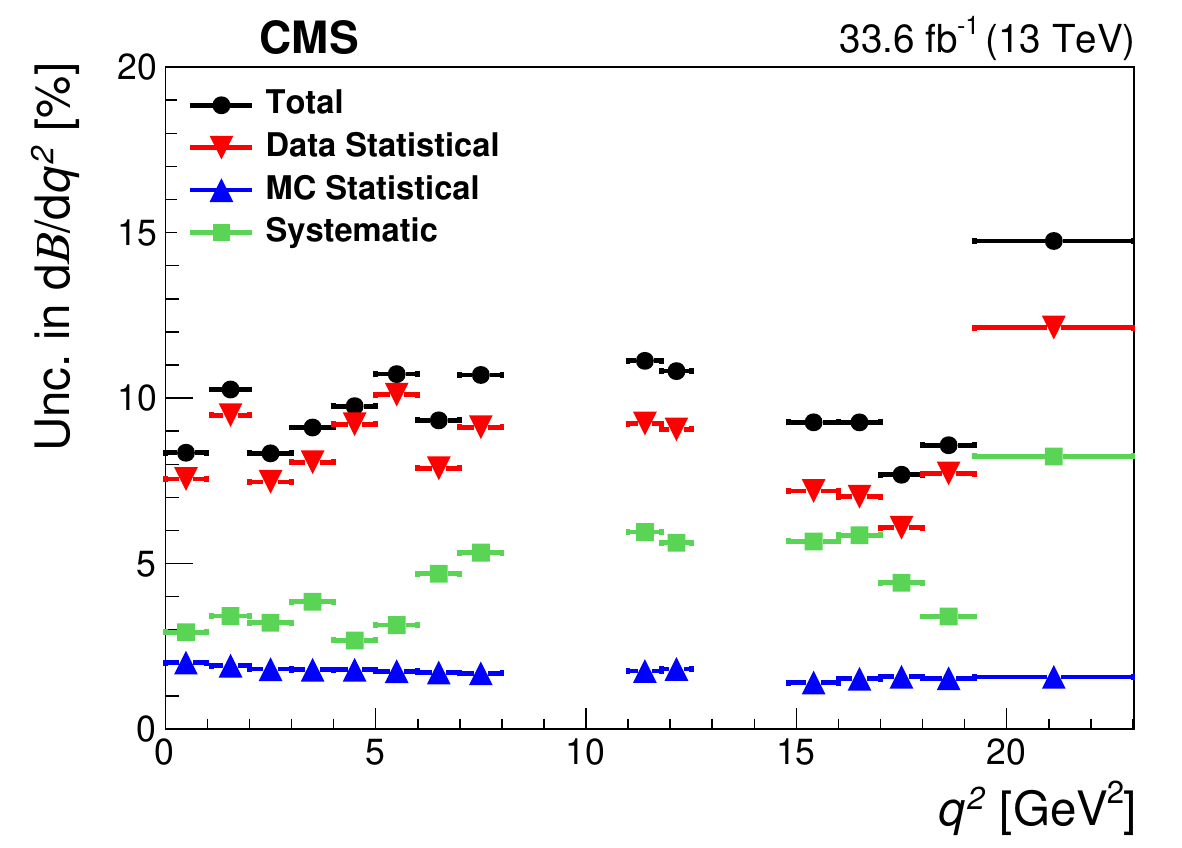}
	\caption{Relative uncertainties in the differential branching fraction measurement of \BKmm per \qsq bin. Different colors correspond to data statistical, simulation statistical, and systematic uncertainties. 
	\label{fig:dif_syst}}
\end{figure}
\begin{figure*}[p]
	\centering
		\includegraphics[width=0.43\textwidth]{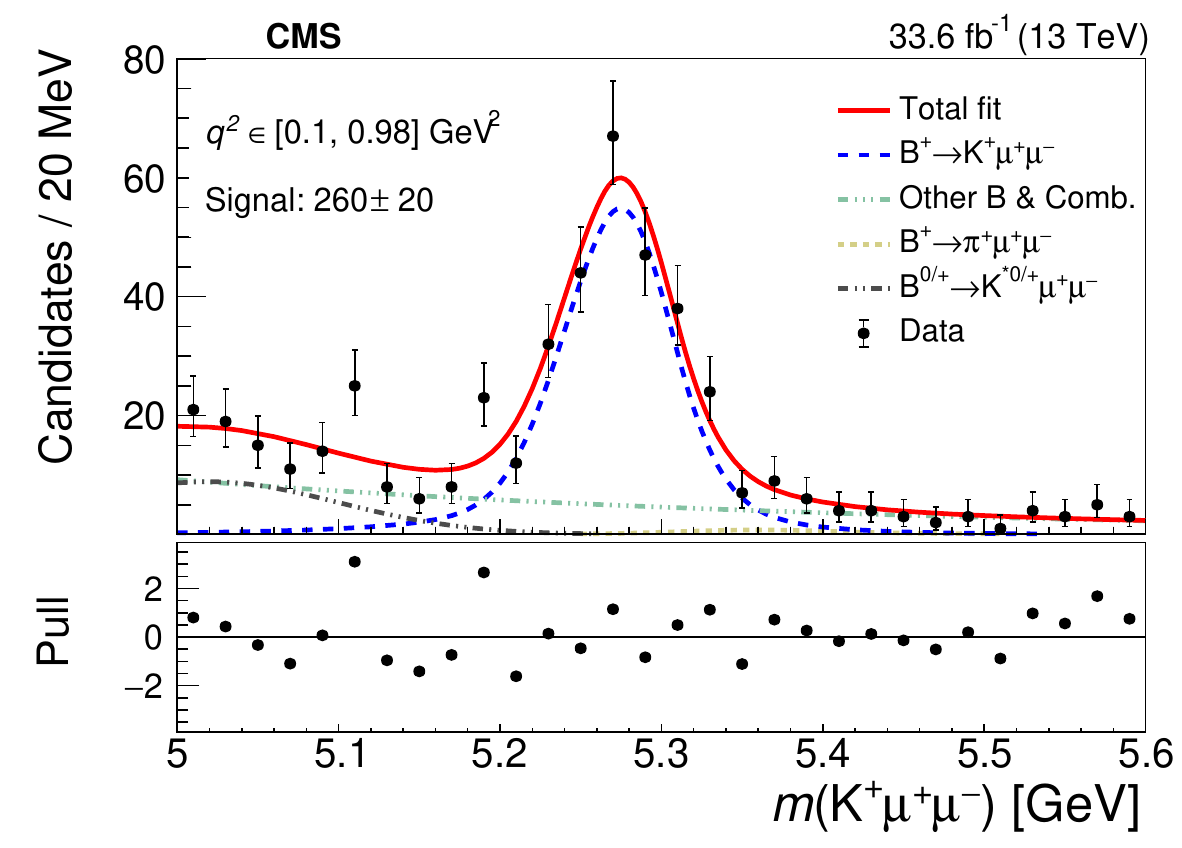} 
		\includegraphics[width=0.43\textwidth]{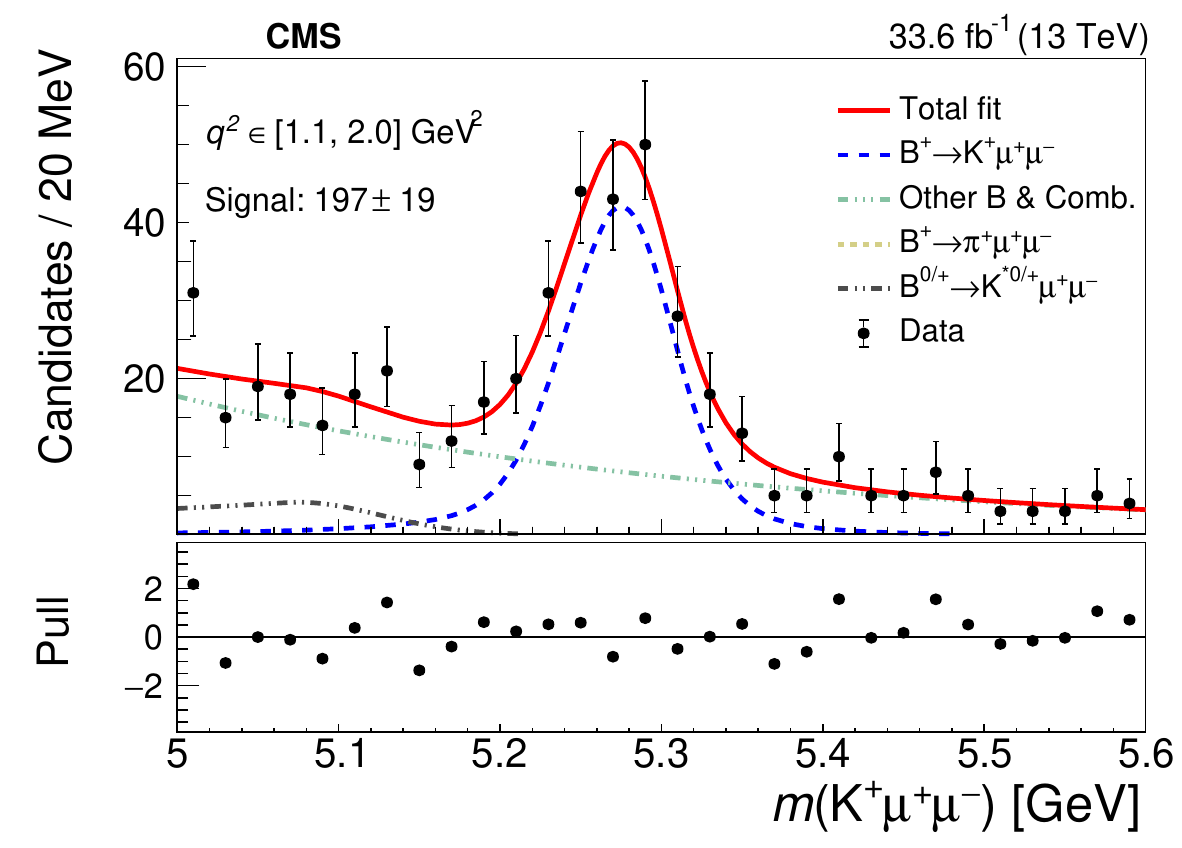} 
		\includegraphics[width=0.43\textwidth]{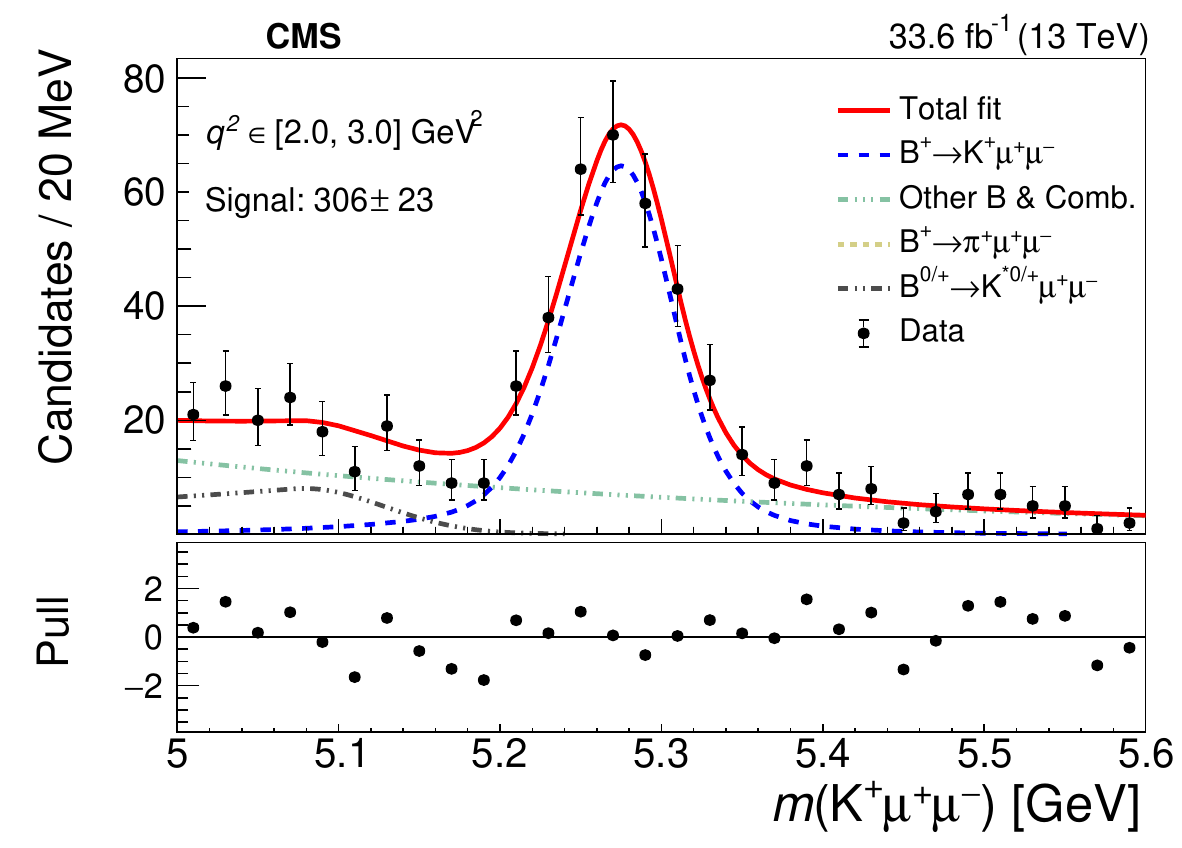} 
		\includegraphics[width=0.43\textwidth]{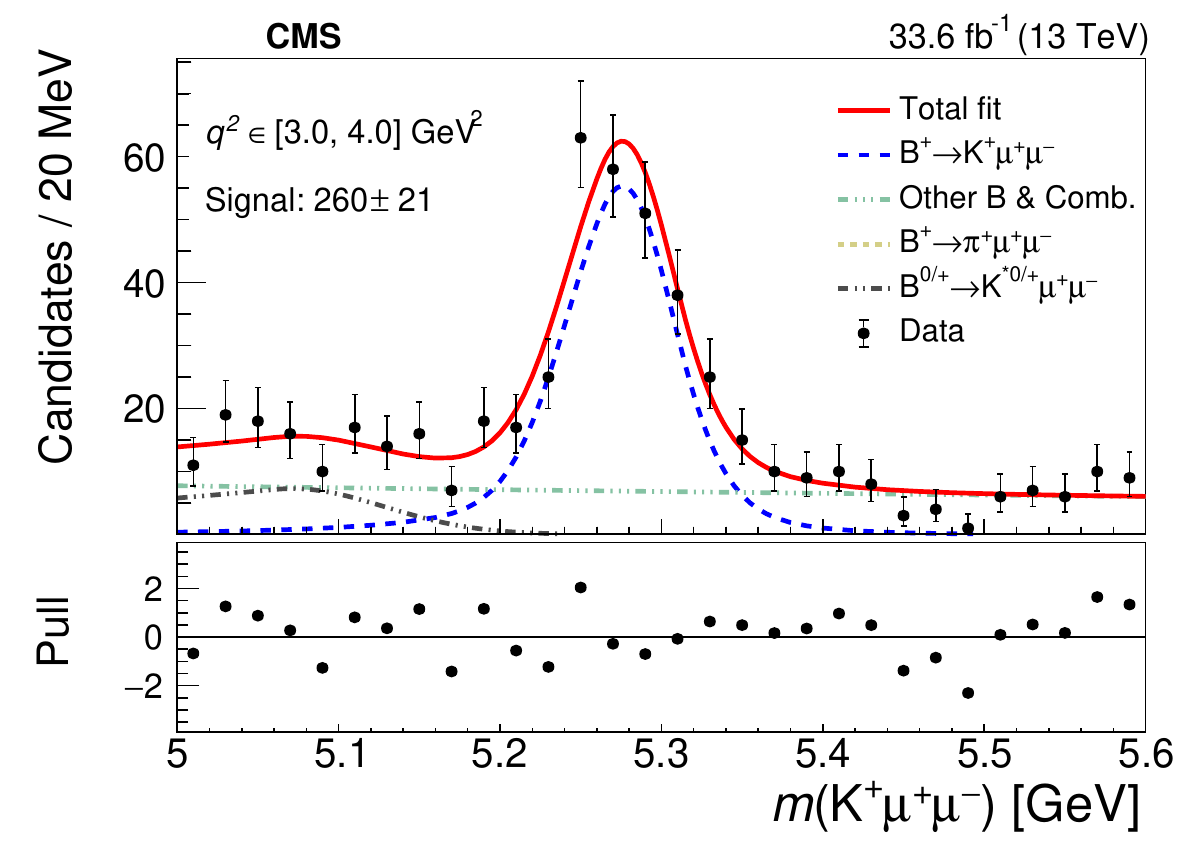}
		\includegraphics[width=0.43\textwidth]{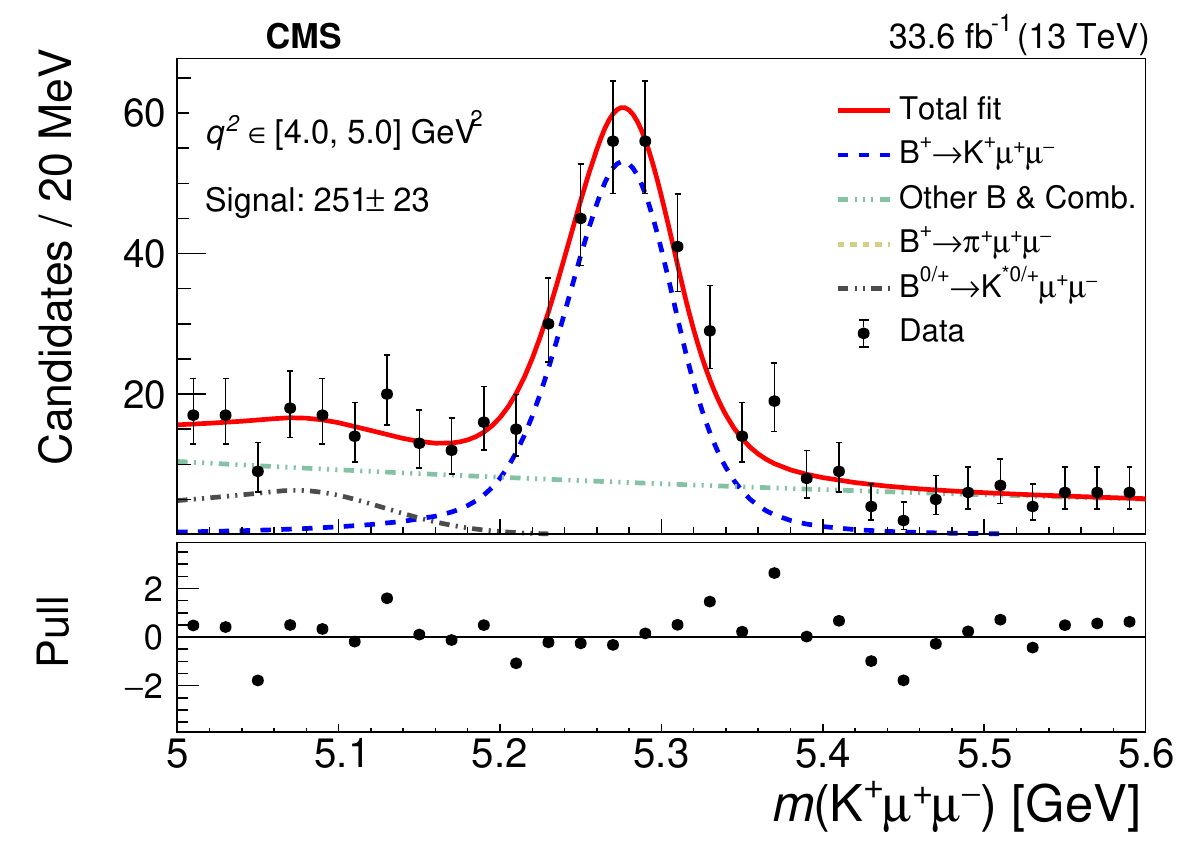}			
		\includegraphics[width=0.43\textwidth]{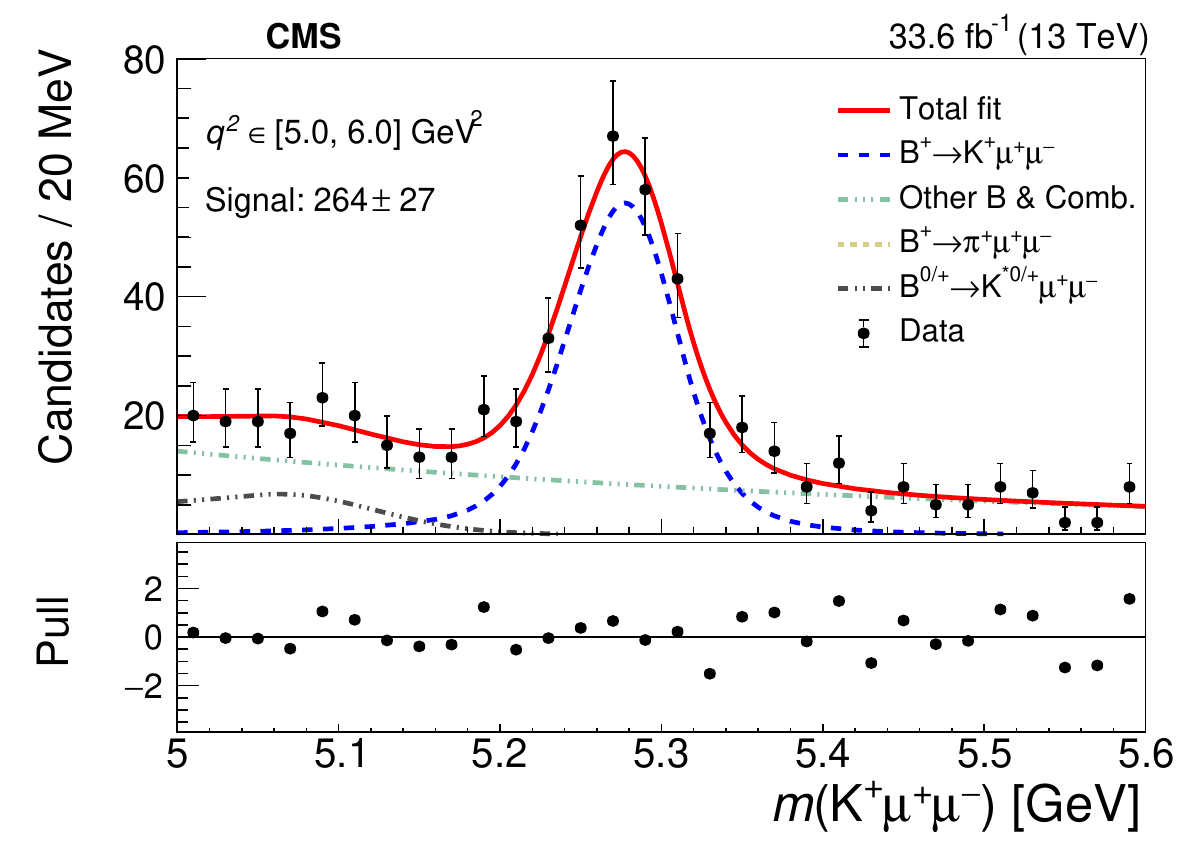} 
		\includegraphics[width=0.43\textwidth]{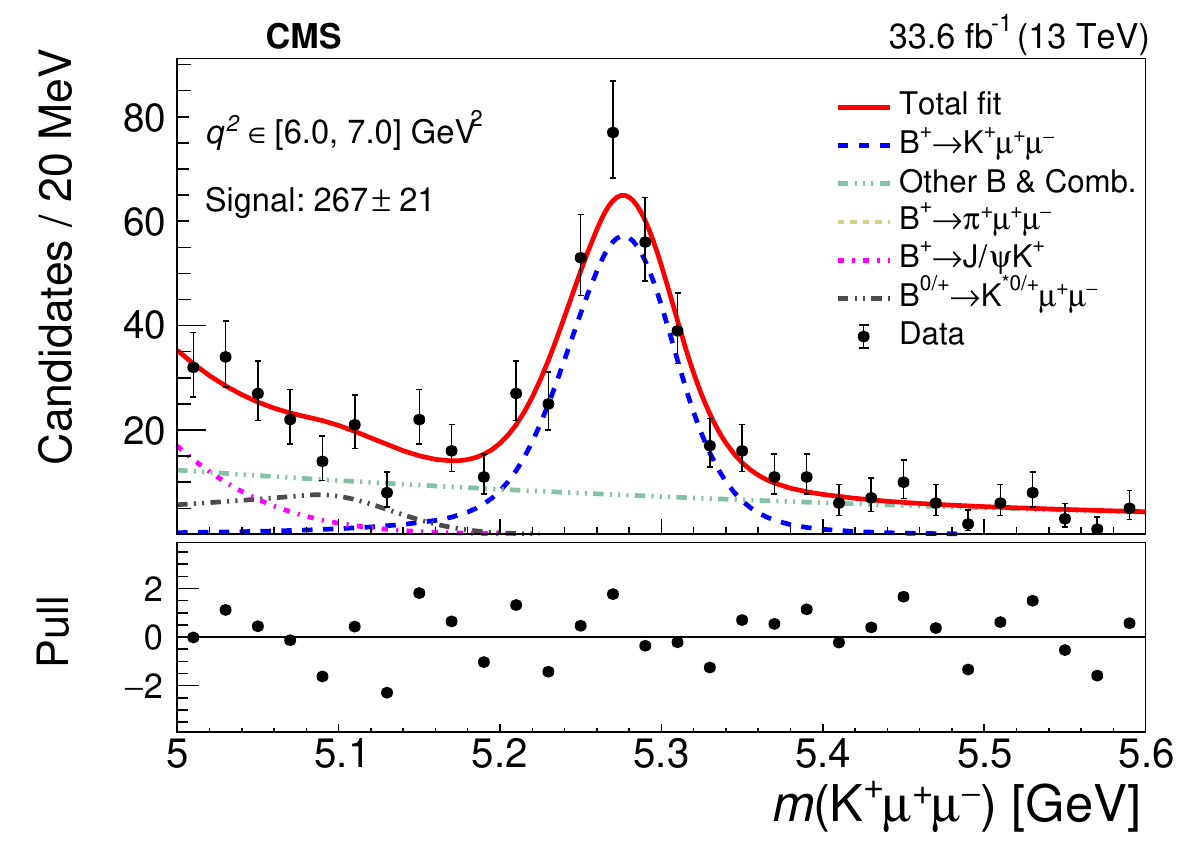}
		\includegraphics[width=0.43\textwidth]{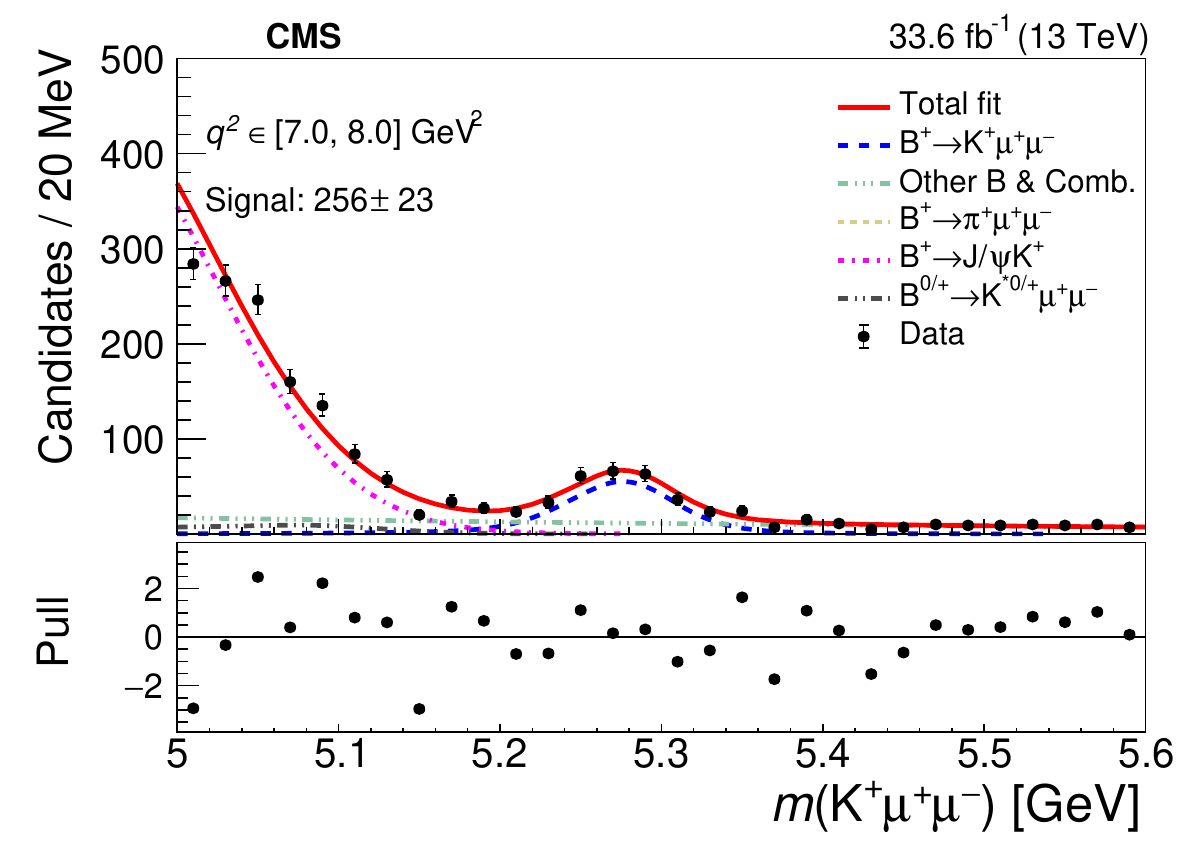}		
  \caption{The \PKp\MM invariant mass distributions in various \qsq bins, with the result of the simultaneous fit overlaid in blue and the individual fit components as described in the legends for (from upper left to lower right): $[0,0.98]$, $[1.1,2.0]$, $[2.0,3.0]$ $[3.0,4.0]$, $[4.0,5.0]$, $[5.0,6.0]$, $[6.0,7.0]$, and $[7.0,8.0]$, \qsq bins. Notations are as in Fig.~\protect\ref{fig:fit_muons}.}
	\label{fig:dif_fit1}
\end{figure*}
\begin{figure*}[p]
	\centering
		\includegraphics[width=0.43\textwidth]{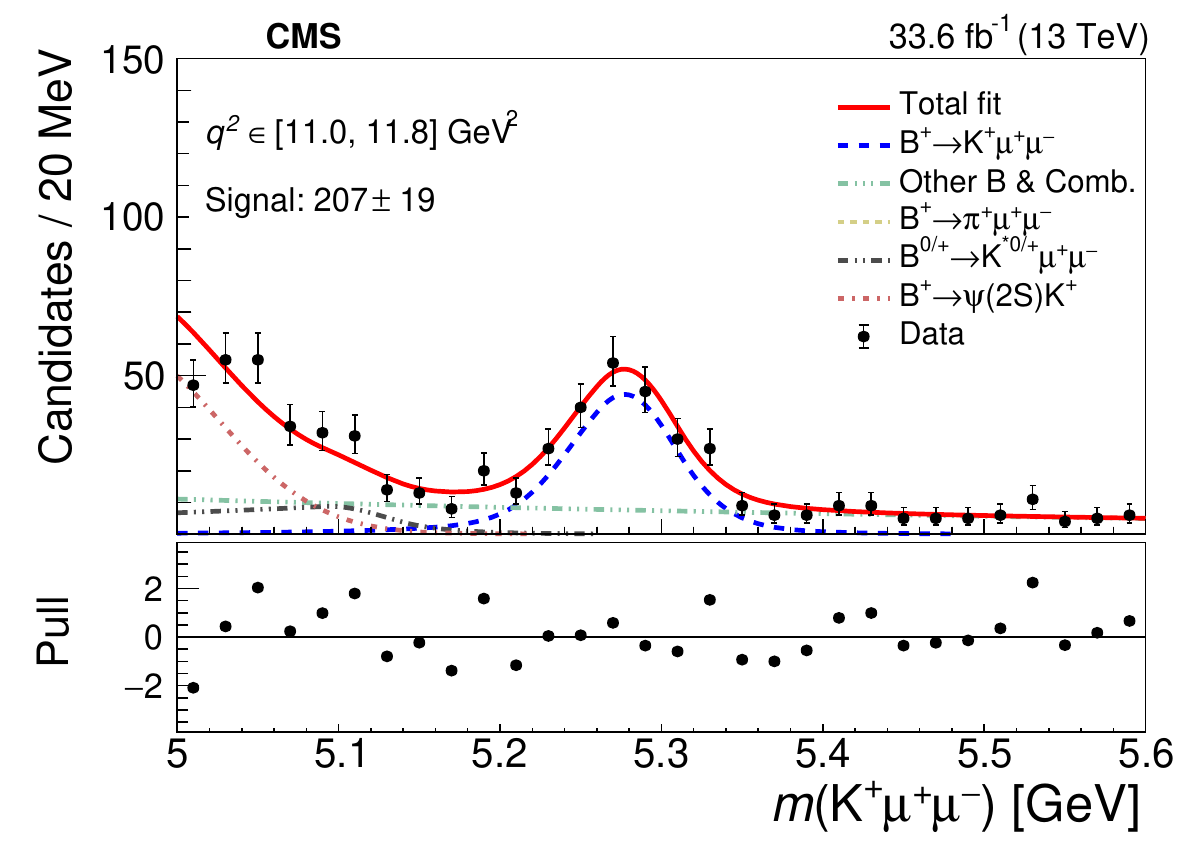}
		\includegraphics[width=0.43\textwidth]{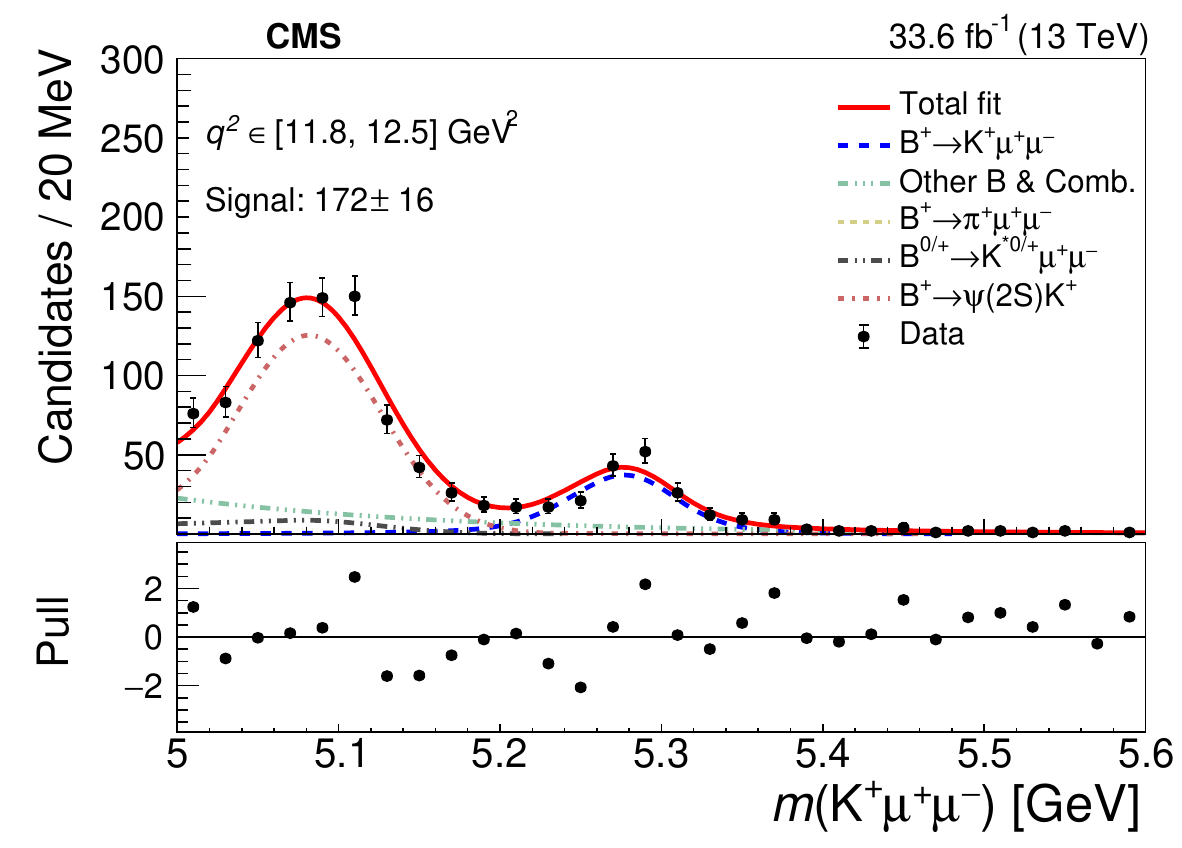}
		 \includegraphics[width=0.43\textwidth]{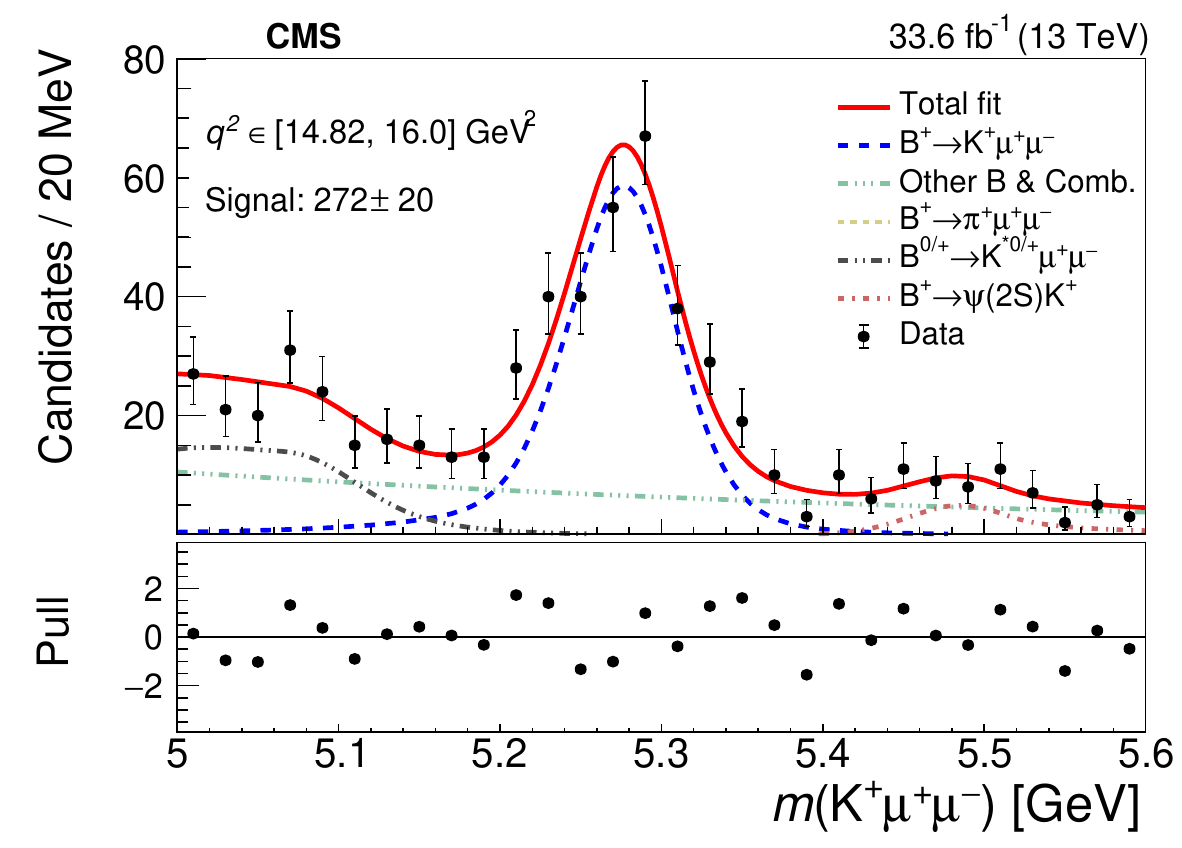}
		 \includegraphics[width=0.43\textwidth]{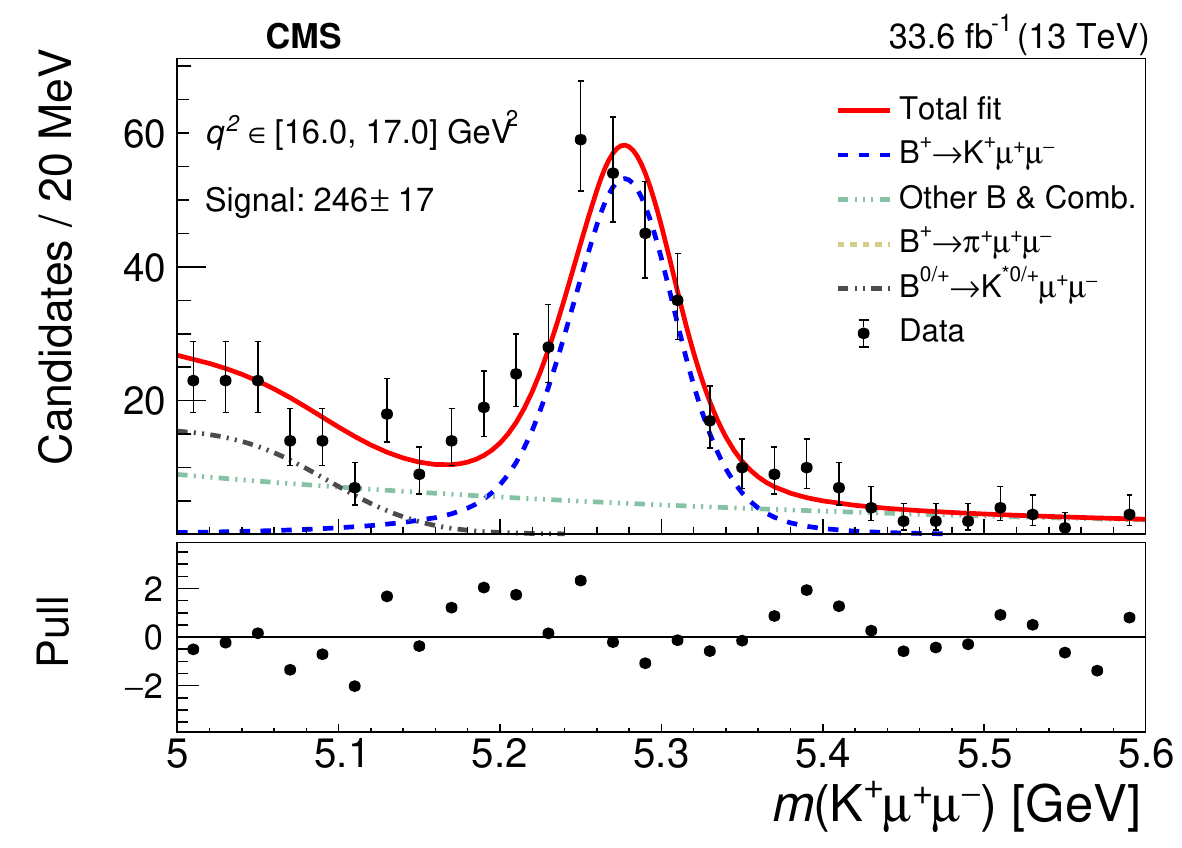} 
		\includegraphics[width=0.43\textwidth]{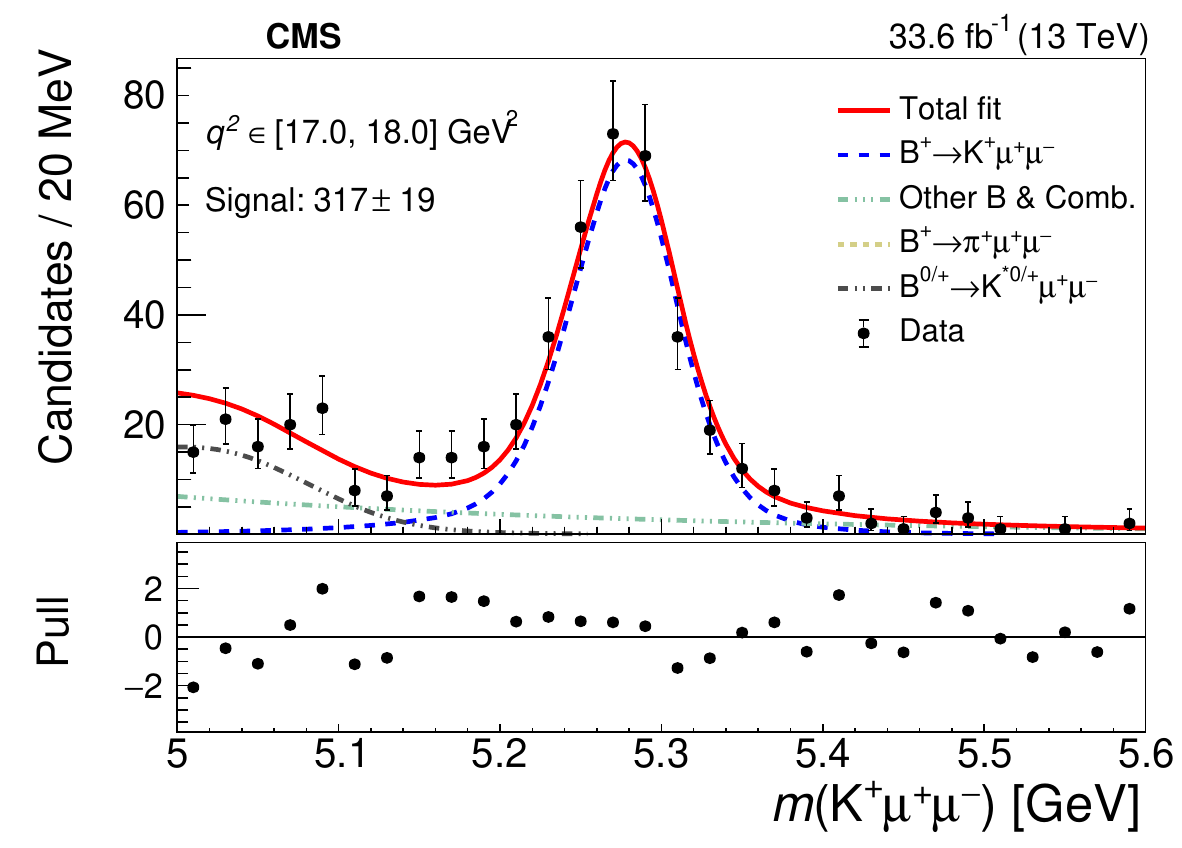}
		\includegraphics[width=0.43\textwidth]{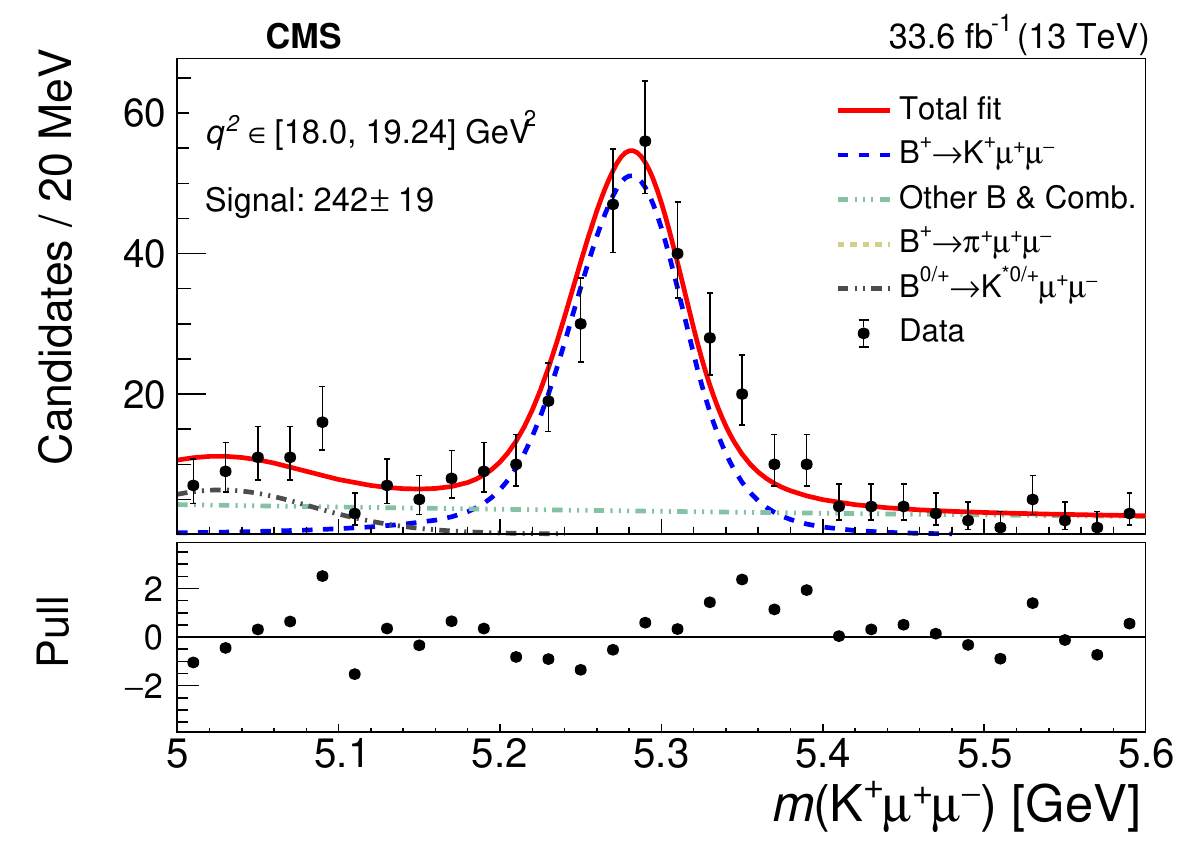}			
		\includegraphics[width=0.43\textwidth]{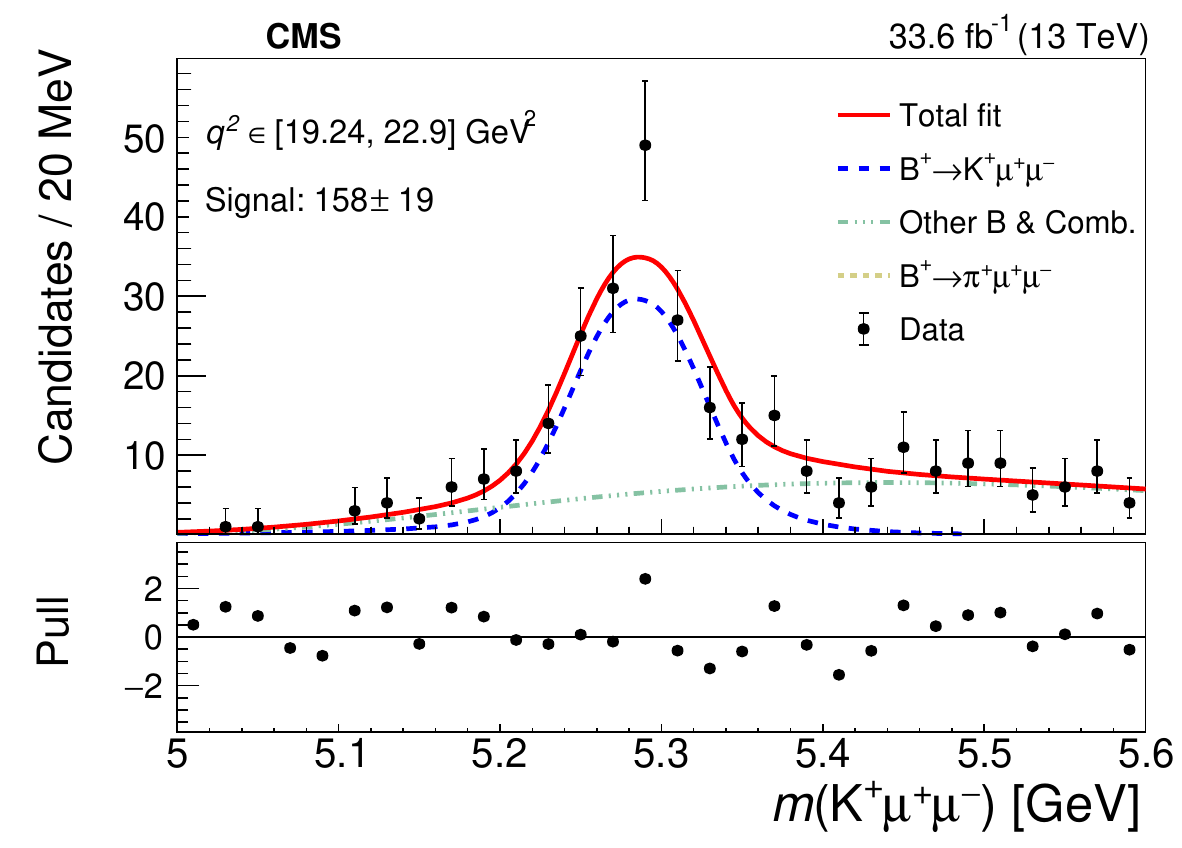}\hfill
	\caption{The \PKp\MM invariant mass distributions in various \qsq bins, with the result of the simultaneous fit overlaid in blue and the individual fit components as described in the legends for (from upper left to lower right):  $[11.0,11.8]$, $[11.8,12.5]$, $[14.82,16.0]$, $[16.0,17.0]$, $[17.0,18.0]$, $[18.0,19.24]$, and $[19.24,22.9]\GeV^{2}$ \qsq bins. Notations are as in Fig.~\protect\ref{fig:fit_muons}.}
\label{fig:dif_fit2}
\end{figure*}
\begin{figure*}[htb]
        \centering
        \includegraphics[width=0.8\textwidth]{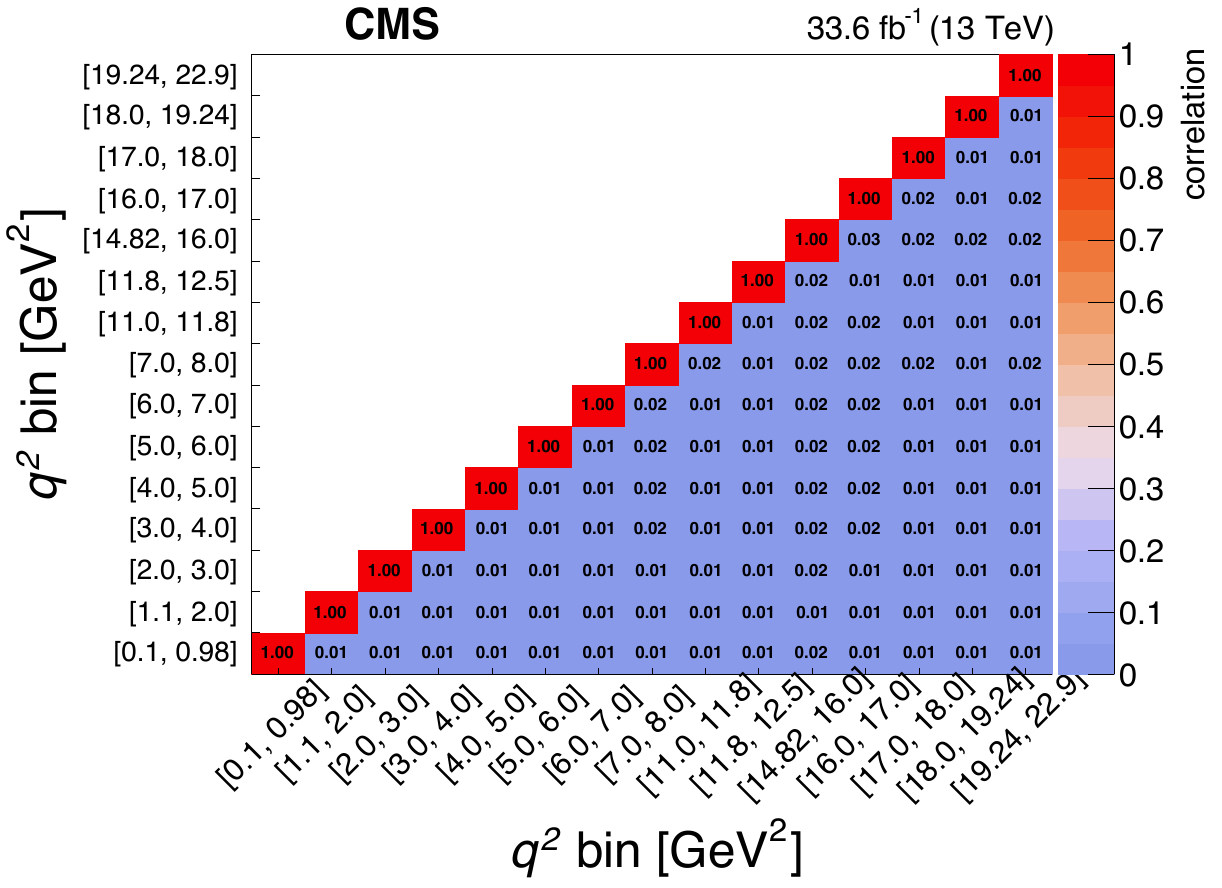}
        \caption{Correlation matrix for the differential branching fraction extraction between different \qsq bins in the simultaneous fit.
        \label{fig:dif_cor}}
\end{figure*}
\begin{figure*}[htb]
        \centering
        \includegraphics[width=0.8\textwidth]{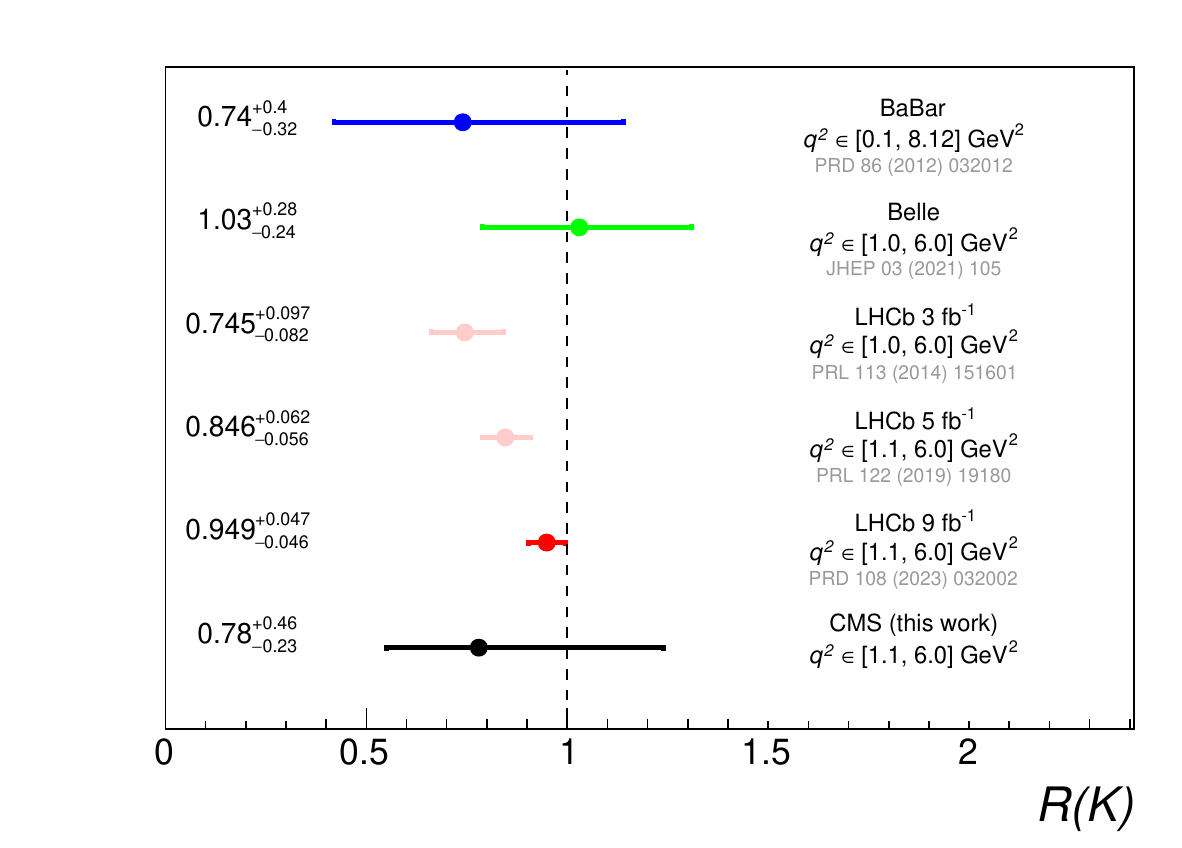}
        \caption{Summary of \RK measurements from BaBar~\cite{BaBar:2012mrf}, Belle~\cite{BELLE:2019xld}, and LHCb~\cite{LHCb:2014vgu,LHCb:2019hip,LHCb:2021trn} experiments, as well as the present CMS measurement. The pink data points of the first three LHCb measurements were superseded by the latest one, shown as the red point.
        \label{fig:summary}}
\end{figure*}
\clearpage

\subsection{\texorpdfstring{${\boldmath R(\PK)}$}{R(K)} measurement formalism}
\label{sec:formalism}
The experimentally accessible equivalent of Eq.~(\ref{eq:RKJ}) can be written as follows, using the event yields $N$ from the fits to the \PB candidate mass spectra and the products of acceptances and efficiencies  $(\Aepst$ in the muon channel and \Aeps in the electron channel):
\begin{linenomath}
\begin{multline}
		\RK =\left(\frac{N_{\BKmm}/N_{\BKJpmm}}{N_{\BKee}/N_{\BKJpee}}\right)\\
		\times\left(\frac{(\Aeps)_{\BKee}/(\Aeps)_{\BKJpee}}{(\Aepst)_{\BKmm}/(\Aepst)_{\BKJpmm}}\right).
\end{multline}
\end{linenomath}

In order to obtain the best fit value of \RK and its confidence interval, the measured quantities $N_{\BKmm}$ and $N_{\BKee}$ are expressed as
\begin{linenomath}
\ifthenelse{\boolean{cms@external}}
{ 
\begin{multline}
  N_{\BKmm} = \RBR\frac{N_{\BKJpmm}}{\BR(\JPsi \!\to\! \MM)} \\
  \times \frac{(\Aepst)_{\BKmm}}{(\Aepst)_{\BKJpmm}}
    \label{eq:rk_expression_muon}
\end{multline}
} 
{ 
\begin{equation}
    N_{\BKmm} = \RBR\frac{N_{\BKJpmm}}{\BR(\JPsi \!\to\! \MM)}\frac{(\Aepst)_{\BKmm}}{(\Aepst)_{\BKJpmm}}
    \label{eq:rk_expression_muon}
\end{equation}
}
\end{linenomath}
and
\begin{linenomath}
\ifthenelse{\boolean{cms@external}}
{ 
\begin{multline}
  N_{\BKee}=\iRK \RBR \frac{N_{\BKJpee}}{\BR(\JPsi\!\to\! \EE)} \\
  \times \frac{(\Aeps)_{\BKee}}{(\Aeps)_{\BKJpee}},
    \label{eq:rk_expression_electron}
\end{multline}
} 
{ 
\begin{equation}
    N_{\BKee}=\iRK \RBR \frac{N_{\BKJpee}}{\BR(\JPsi\!\to\! \EE)} \cdot \frac{(\Aeps)_{\BKee}}{(\Aeps)_{\BKJpee}},
    \label{eq:rk_expression_electron}
\end{equation}
}
\end{linenomath}
where $\RBR = \frac{\BR\left(\BKmm \right)}{\BR\left(\PBp \!\to\! \PKp\JPsi\right)}$. We note that \iRK is used as the parameter of interest instead of \RK because this choice makes the likelihood significantly more Gaussian.

A simultaneous fit is done using Eqs. (\ref{eq:rk_expression_muon}, \ref{eq:rk_expression_electron}), with the likelihood function in each lepton channel defined as
\begin{linenomath}
\ifthenelse{\boolean{cms@external}}
{ 
\begin{multline}
		\mathcal{L}^{\PGm} = \mathcal{L}_{\text{low-\qsq}}^{\PGm}(N_{\BKmm})  \mathcal{L}_{\JPsi}^\PGm(N_{\BKJpmm})\\
		\times  G((\Aepst)_{\BKmm}) \\
                \times G((\Aepst)_{\BKJpmm}) G(\BR_{\JPsi}^{\PGm}),
	\label{eq:R_Km}
\end{multline}
\begin{multline}
		\mathcal{L}^{\Pe} = \mathcal{L}_{\text{low-\qsq}}^{\Pe}(N_{\BKee})  \mathcal{L}_{\JPsi}^\Pe(N_{\BKJpee})\\
		\times  G((\Aeps)_{\BKee}) \\
                \times G((\Aeps)_{\BKJpee}) G(\BR_{\JPsi}^{\Pe}),
	\label{eq:R_Ke}
\end{multline}
} 
{ 
\begin{multline}
		\mathcal{L}^{\PGm} = \mathcal{L}_{\text{low-\qsq}}^{\PGm}(N_{\BKmm})  \mathcal{L}_{\JPsi}^\PGm(N_{\BKJpmm})\\
		\times  G((\Aepst)_{\BKmm}) G((\Aepst)_{\BKJpmm}) G(\BR_{\JPsi}^{\PGm}),
	\label{eq:R_Km}
\end{multline}
\begin{multline}
		\mathcal{L}^{\Pe} = \mathcal{L}_{\text{low-\qsq}}^{\Pe}(N_{\BKee})  \mathcal{L}_{\JPsi}^\Pe(N_{\BKJpee})\\
		\times  G((\Aeps)_{\BKee}) G((\Aeps)_{\BKJpee}) G(\BR_{\JPsi}^{\Pe}),
	\label{eq:R_Ke}
\end{multline}
}
\end{linenomath}
where $\mathcal{L}_{\text{low-\qsq}}^\ell$ and $\mathcal{L}_{\JPsi}^{\ell}$ ($\ell = \PGm$ or $\Pe$) are the mass fit likelihood function used in the low-\qsq and \JPsi regions, respectively,  $G$ is the Gaussian function with the mean at the nominal value of the argument and the RMS given by the corresponding uncertainty, and $\BR_{\JPsi}^{\ell}$ is the branching fraction of the $\JPsi \!\to\! \ell\ell$ decay. The value of \RK is measured by maximizing the likelihood function $\mathcal{L} = \mathcal{L}^\PGm\mathcal{L}^\Pe$, either for a single dielectron category or for the two categories simultaneously, which is achieved by taking $\mathcal{L}^\Pe = \mathcal{L}^\text{PF-PF}\mathcal{L}^\text{PF-LP}$. 
\cleardoublepage \section{The CMS Collaboration \label{app:collab}}\begin{sloppypar}\hyphenpenalty=5000\widowpenalty=500\clubpenalty=5000
\cmsinstitute{Yerevan Physics Institute, Yerevan, Armenia}
{\tolerance=6000
A.~Hayrapetyan, A.~Tumasyan\cmsAuthorMark{1}\cmsorcid{0009-0000-0684-6742}
\par}
\cmsinstitute{Institut f\"{u}r Hochenergiephysik, Vienna, Austria}
{\tolerance=6000
W.~Adam\cmsorcid{0000-0001-9099-4341}, J.W.~Andrejkovic, T.~Bergauer\cmsorcid{0000-0002-5786-0293}, S.~Chatterjee\cmsorcid{0000-0003-2660-0349}, K.~Damanakis\cmsorcid{0000-0001-5389-2872}, M.~Dragicevic\cmsorcid{0000-0003-1967-6783}, P.S.~Hussain\cmsorcid{0000-0002-4825-5278}, M.~Jeitler\cmsAuthorMark{2}\cmsorcid{0000-0002-5141-9560}, N.~Krammer\cmsorcid{0000-0002-0548-0985}, A.~Li\cmsorcid{0000-0002-4547-116X}, D.~Liko\cmsorcid{0000-0002-3380-473X}, I.~Mikulec\cmsorcid{0000-0003-0385-2746}, J.~Schieck\cmsAuthorMark{2}\cmsorcid{0000-0002-1058-8093}, R.~Sch\"{o}fbeck\cmsorcid{0000-0002-2332-8784}, D.~Schwarz\cmsorcid{0000-0002-3821-7331}, M.~Sonawane\cmsorcid{0000-0003-0510-7010}, S.~Templ\cmsorcid{0000-0003-3137-5692}, W.~Waltenberger\cmsorcid{0000-0002-6215-7228}, C.-E.~Wulz\cmsAuthorMark{2}\cmsorcid{0000-0001-9226-5812}
% ...
\par}
\vskip\cmsinstskip
\dag:~Deceased\\
$^{1}$Also at Yerevan State University, Yerevan, Armenia\\
$^{2}$Also at TU Wien, Vienna, Austria\\
% ...
\end{sloppypar}
\end{document}